\documentclass{article}

\usepackage{PRIMEarxiv}

\usepackage[utf8]{inputenc} 
\usepackage[T1]{fontenc}    
\usepackage[hidelinks]{hyperref}       
\usepackage{url}            
\usepackage{booktabs}       
\usepackage{amsfonts}       
\usepackage{nicefrac}       
\usepackage{microtype}      
\usepackage{lipsum}
\usepackage{fancyhdr}       
\usepackage{graphicx}       

\usepackage{float}
\usepackage{amsmath,amssymb}
\usepackage[ruled,vlined]{algorithm2e}
\SetKwInOut{Input}{Input}
\SetKwInOut{Output}{Output}
\SetKwRepeat{Do}{do}{while}

\usepackage{grffile}

\usepackage{bbm}

\usepackage{amssymb}
\usepackage{amsmath,amsfonts,amsthm,amssymb,dsfont}

\usepackage[table, svgnames, dvipsnames]{xcolor}
\usepackage{longtable}
\usepackage{makecell, cellspace, caption}

\title{Ultrafast learning of 4-node hybridization cycles in phylogenetic networks using algebraic invariants}

\author{ {Zhaoxing Wu} \\
	Wisconsin Institute for Discovery\\
	Department of Statistics\\
	University of Wisconsin-Madison\\
	Madison, WI 53706 \\
	\And
	{Claudia Sol\'is-Lemus}\thanks{Corresponding author: solislemus@wisc.edu} \\
	Wisconsin Institute for Discovery\\
	Department of Plant Pathology\\
	University of Wisconsin-Madison\\
	Madison, WI 53706 \\
}

\hypersetup{
pdftitle={Phylogenetic invariants},
pdfsubject={q-bio.NC, q-bio.QM},
pdfauthor={Bella Wu, Claudia Solis-Lemus},
pdfkeywords={First keyword, Second keyword, More},
}

\begin{document}
\maketitle

\begin{abstract}
\textbf{Motivation:} The abundance of gene flow in the Tree of Life challenges the notion that evolution can be represented with a fully bifurcating process, as this process cannot capture important biological realities like hybridization, introgression, or horizontal gene transfer. 
Coalescent-based network methods are increasingly popular, yet not scalable for big data, because they need to perform a heuristic search in the space of networks as well as numerical optimization that can be NP-hard. 

\textbf{Results:} Here, we introduce a novel method to reconstruct phylogenetic networks based on algebraic invariants. While there is a long tradition of using algebraic invariants in phylogenetics, our work is the first to define phylogenetic invariants on concordance factors (frequencies of 4-taxon splits in the input gene trees) to identify level-1 phylogenetic networks under the multispecies coalescent model. Our novel inference methodology is optimization-free as it only requires the evaluation of polynomial equations, and as such, it bypasses the traversal of network space, yielding a computational speed at least 10 times faster than the fastest-to-date network methods.
We illustrate the accuracy and speed of our new method on a variety of simulated scenarios as well as in the estimation of a phylogenetic network for the genus \textit{Canis}.

\textbf{Availability and Implementation:} We implement our novel theory on an open-source publicly available Julia package \texttt{PhyloDiamond.jl} available at \url{https://github.com/solislemuslab/PhyloDiamond.jl} with broad applicability within the evolutionary biology community. 

\textbf{Contact:} solislemus@wisc.edu
\end{abstract}

\keywords{Hybridization \and Polynomials \and Reticulate evolution \and Coalescent model}

\section{Introduction}
The Tree of Life is the graphical structure that represents the evolutionary process from single-cell organisms at the origin of life to present day biodiversity. Mathematically, a phylogenetic tree is a fully bifurcating graph in which its internal nodes represent speciation events that give rise to two children nodes. 
Recent evidence \cite{de2021biological, adavoudi2021consequences, perez2021molecular, rey2021diverging, suvorov2022deep} has challenged the notion that evolution across the Tree of Life can be represented with a fully bifurcating process, as this process cannot capture important biological realities like hybridization, introgression, or horizontal gene transfer that require two fully
separated branches to join again. These processes of reticulate evolution are more prevalent in certain groups like plants \cite{charles2021assisted}, fungi \cite{steensels2021interspecific} and prokaryotes \cite{diop2022gene}. To accurately include these groups in the Tree of Life, recent years have seen an explosion of methods to reconstruct phylogenetic networks, which naturally account for reticulate evolution \cite{Degnan2018, elworth2019advances, blair2020phylogenetic, kong2022classes}. Existing methods, however, are still not scalable enough to tackle the complexities of the present day's biological big data. 
The most scalable alternatives infer split (or implicit) networks \cite{bryant2004neighbor, huson2006application, grunewald2007qnet} which are not biologically interpretable as internal nodes no longer represent biological events (speciation or hybridization), and the resulting network is completely unrooted, even with the inclusion of an outgroup.

Among methods to infer explicit networks (those whose internal nodes do represent speciation or hybridization events), likelihood-based approaches are among the most popular ones. \texttt{PhyloNet} \cite{Yu2014} has been the pioneer of likelihood inference of phylogenetic networks, later expanding to Bayesian \cite{Wen2016} and pseudolikelihood alternatives \cite{yu2015maximum}. However, even in its most scalable option (pseudolikelihood), the method is still not suitable beyond dozens of taxa and hundreds of genes. \texttt{SNaQ} \cite{snaq} within the \texttt{PhyloNetworks} Julia package \cite{Solis-Lemus2017-jk} has proven a scalable alternative to infer large phylogenetic networks from multilocus alignments. This method is based on a pseudolikelihood model under the coalescent, and its running time does not increase as a function of number of genes because the input data for \texttt{SNaQ} are split frequencies on subsets of $4$ taxa (concordance factors). That is, after the estimation of the concordance factors, the running time is the same for data with 10 genes or 10,000 genes. Despite its popularity, \texttt{SNaQ} still lacks scalability as the number of taxa increases and remains a suitable alternative only for $50$ or fewer taxa.
Last, \texttt{BEAST2} \cite{Zhang2018} provides the only co-estimation method that infers simultaneously the phylogenetic networks and the gene trees, and it allows users to infer a variety of relevant biological parameters (such as divergence times), yet the complexity of this method renders it unsuitable for anything beyond a handful of taxa.

Because of the complexity in inferring a phylogenetic network, the evolutionary biology community has embraced the use of hybrid detection methods such as ABBA-BABA test \cite{pattersonD}, \texttt{MSCquartets} \cite{Allman-Singularities-and-Boundaries, MSCquartets-RPackage}, and \texttt{HyDe} \cite{hyde-paper} to identify hybridization events on a fixed phylogeny. These methods take in subsets of taxa and test whether the current subset can be well-explained with a tree-like pattern or if a hybridization event exists in this subset. While fast, these methods can be inaccurate in the presence of multiple hybridization events affecting the same taxa \cite{Bjorner2022-wg} or by ghost lineages \cite{Pang2022-uq, Tricou2022-qv, Bjorner2022-wg}. Furthermore, the process of first reconstructing a phylogenetic tree and then adding hybridization events can be flawed given the bias that gene flow causes on the inference of the backbone phylogenetic tree \cite{Solis-Lemus2016-jr}.

Here, we introduce a novel method to infer hybridization events using phylogenetic invariants. Our method only requires evaluation of polynomial equations, so it bypasses optimization in the space of networks and it is at least $10$ times faster than current inference network methodologies. Using gene trees as input, our method exploits the signal in the frequencies of splits in the data, also well-known as \emph{concordance factors}. Indeed, under the coalescent model, splits of taxa display certain probabilities of appearing in the sample of gene trees. Under the true network, these frequencies need to satisfy certain polynomial equations, denoted \emph{phylogenetic invariants}. By plugging in the observed frequencies on the phylogenetic invariants of every possible placement of the hybridization cycle, we can identify the hybridization cycle that better agrees with the data as the one whose evaluated invariants are closest to zero. 

Algebraic invariants have been widely used in phylogenetics to identify trees under a variety of models of evolution (Jukes-Cantor \cite{felsenstein1991counting, steel1995classifying}, GTR \cite{allman2003phylogenetic, Fernandez-Sanchez2016-yr, Casanellas2021-ew, Casanellas2023-yn}), and more recently, to identify phylogenetic networks \cite{Gross2018, gross2021distinguishing, ardiyansyah2021distinguishing, cummings2021invariants, Barton2022}.
Nevertheless, existing work on phylogenetic invariants on networks has been restricted to models of evolution. Our work is the first to define phylogenetic invariants on concordance factors under the multispecies coalescent model on networks.

This work first introduces the phylogenetic invariants that concordance factors need to satisfy under level-1 phylogenetic networks with a 4-node hybridization cycle under the multispecies coalescent model. Next, we describe an inference methodology to use the phylogenetic invariants to infer the correct 4-node hybridization cycle among $n$ taxa.
Moreover, we demonstrate the performance of our method with simulated data, and we revisit the inference of the \emph{Canis} phylogenetic network that had already been published in \cite{dog}. We show that our method is faster than all existing network inference methods, and it accurately identifies the correct hybridization cycle in the \emph{Canis} genus. Additionally, in all simulating scenarios, our method recognizes the correct placement of hybridization events with high accuracy. Even in the few cases when our method identifies an incorrect network (usually cases when there is not sufficient sampling of taxa that descend from the hybrid node), the correct network is still ranked among top 5 optimal networks given the invariant score (evaluated phylogenetic invariants that need to vanish to zero). This means that, regardless of sampling, our method is always able to reduce the space of candidate networks that can later be tested with a likelihood-based approach.

We highlight that our method has two main limitations: 1) it is only suitable to identify level-1 hybridization cycles with 4 nodes, and 2) it sometimes fails to identify the correct network as the one with the top 1 ranked invariant score when there is only one taxon sampled that is descendant of the hybrid node. Despite these limitations, our method is a valuable innovation in the landscape of phylogenetic network methods, if anything, to reduce the space of candidate networks to be used in likelihood-based methodologies. Indeed, we do not consider our method as a replacement for other network inference approaches. On the contrary, we believe that our ultrafast learning methodology can serve to 1) identify the true phylogenetic network when it involves a simple hybridization event, and 2) provide several candidate networks to be later tested with a likelihood-based approach, effectively bypassing heuristic optimization in the space of networks.
Furthermore, our method is open source, publicly available as the new Julia package \texttt{PhyloDiamond.jl} in \url{https://github.com/solislemuslab/PhyloDiamond.jl}.

The structure of the paper is as follows. In \nameref{methods}, we introduce the phylogenetic invariants under the multispecies coalescent model on networks, and we describe an inference methodology that uses the phylogenetic invariants to identify the 4-node hybridization cycle that generates the data. In this section, we also depict the simulation study and the re-analysis of the \emph{Canis} data from \cite{dog}.
In \nameref{results}, we present the results on the simulated data and the \emph{Canis} phylogeny. Last, in \nameref{sec:disc}, we explain the main limitations of our method and potential future work.

\section{Materials and Methods}
\label{methods}

\subsection{Phylogenetic invariants for 4-node hybridization cycles in level-1 phylogenetic networks}

Under the coalescent model, the distribution of gene trees estimated from multilocus sequence alignments provides information on the true network that generated the data \cite{Yu2014}. In \cite{snaq}, it was shown that the split frequencies on subsets of 4 taxa, namely the \emph{concordance factors}(CF), also provide information on the true network.
That is, a CF of a given quartet (or split) is the proportion of genes whose true tree displays that quartet (or split) \cite{Baum2007}. For example, for a taxon set $s = \{a, b, c, d\}$, there are only three possible quartets, represented by the splits $q_1 = ab|cd$, $q_2 = ac|bd$ and $q_3 = ad|bc$. The CF for the split $ab|cd$ is the proportion of gene trees that display this split.

\begin{figure}[h]
    \centering
    \includegraphics[scale=0.8]{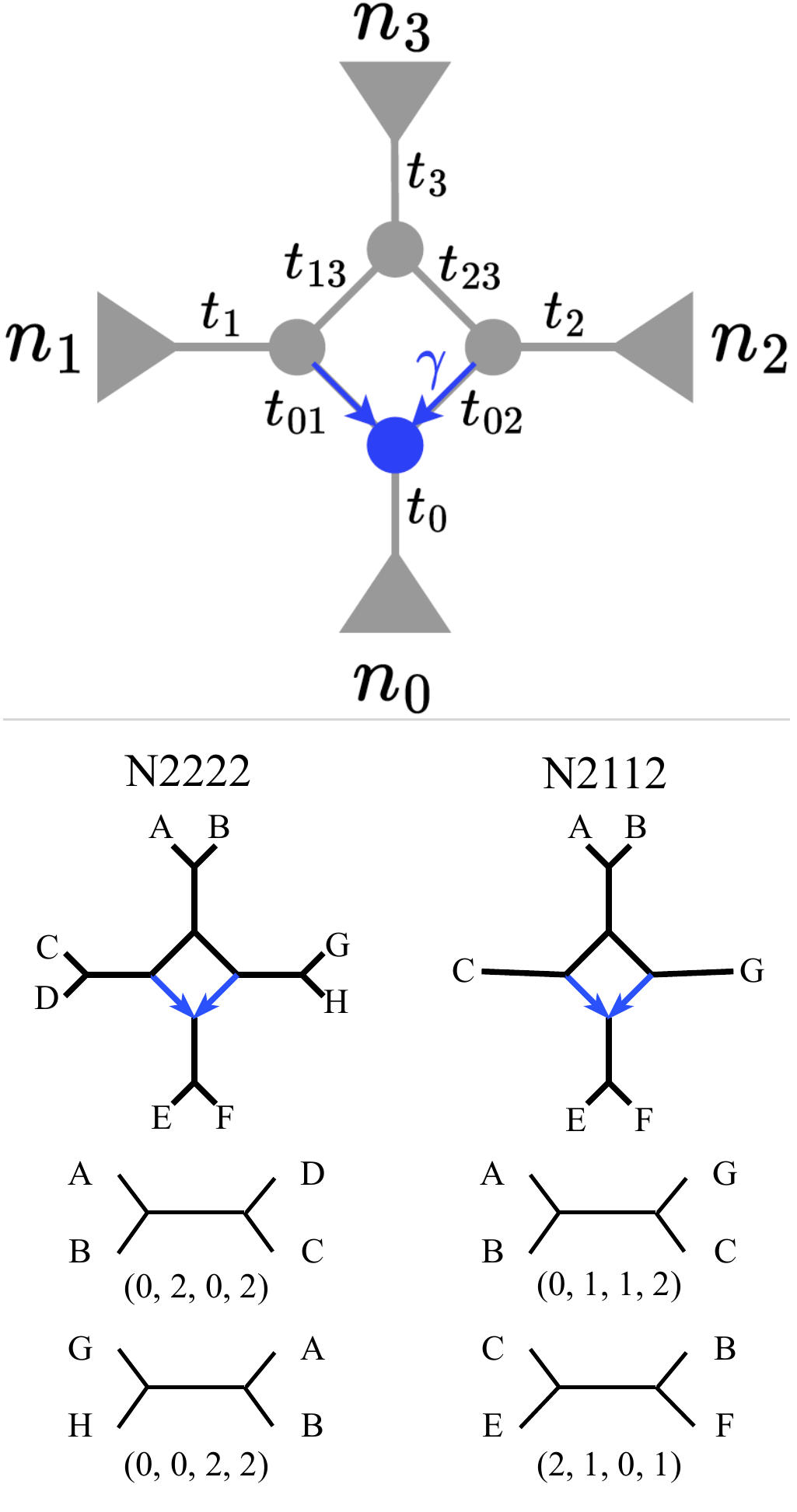}
    \caption{\textbf{Top}: 4-node hybridization cycle on a level-1 phylogenetic network where $t_i$ represents the branch length in coalescent units. The hybrid node and hybrid edges are in blue. The minor hybrid edge is labelled with $\gamma$ that represents the inheritance probability (or proportion of genes transferred through the minor edge). Denote this level-1 semi-directed network with 4-node hybridization cycles as $N$. \textbf{Middle}: Two examples of level-1 networks, each with one 4-node hybridization cycle, and their corresponding notation (N2222 for the 8-taxon network and N2112 for the 6-taxon network). \textbf{Bottom}: Examples of quartets drawn from each network in top right along with their vector notations. For example, the quartet $AB|CD$ drawn from the N2222 network corresponds to the 4-taxon subset $(0,2,0,2)$ where each entry in this vector corresponds to the number of taxa drawn from each of the four clades: $n_0,n_1,n_2,n_3$.}
    \label{fig:4net}
\end{figure}

As in \cite{snaq}, our method uses the CFs as input data, and 
we focus on the case of 4-node hybridization cycles on level-1 phylogenetic networks as shown in Figure \ref{fig:4net} (top).  A level-1 phylogenetic network is a network with no vertex belonging to more than one hybridization cycle. The semi-directed network in Figure \ref{fig:4net} (top) is unrooted, yet the direction of the hybrid edges (in blue) is known so that the root placement of the network is constrained. Namely, the root cannot be anywhere below the hybrid node (in blue). The hybridization cycle partitions the taxa into 4 subsets with $n_0, n_1, n_2$ and $n_3$ as the number of taxa in each subset. For example, the clade below the hybrid node in blue in Figure \ref{fig:4net} (top), also known as the hybrid clade, has $n_0$ taxa, and the two sister clades to the hybrid clade have $n_1$ and $n_2$ taxa. The notation $n_i$ not only represents the number of taxa in a specific clade, this terminology is also used to refer to that specific clade. In fact, we refer to the hybrid clade as the $n_0$ clade. This $n_i$ notation will later help us represent different network structures and 4-taxon subsets. Moreover, each clade in the network could be a subtree or a subnetwork as long as the overall network is level-1 \cite{Huson2010}. Last, we highlight that by 'taxa', we mean either different species or populations, or multiple individuals sampled from the same species.

In this work, we introduce a method to infer 4-node hybridization cycles in level-1 semi-directed networks.
Since we focus on level-1 4-node hybridization cycles (as in Figure \ref{fig:4net} (top)), we utilize a simplifying notation to represent each network. Let $N$ represent an $n-$taxon semi-directed level-1 phylogenetic network, and we represent this network by the number of taxa in its four clades. For example, in Figure \ref{fig:4net}, the network denoted as N2222 has two taxa in each of the four clades, while the network N2112 contains two taxa in clades $n_0$ and $n_3$, and one taxon in clades $n_1$ and $n_2$. 

Another relevant notation is the vector representation of every 4-taxon subset. For example, in Figure \ref{fig:4net}, the 4-taxon subset $(0,2,0,2)$ drawn from network N2222 corresponds to the case when two taxa are taken from $n_1$ and $n_3$ and no taxon is taken from $n_0$ and $n_2$, and this subset corresponds to the quartet $AB|CD$. Another example is the 4-taxon subset $(0,1,1,2)$ drawn from network N2112. Note that since this network only has one taxon in clades $n_1$ and $n_2$, any 4-taxon subset drawn from this network must have at most one taxon in the second and third value in the 4-taxon vector notation. In general, a 4-taxon subset is represented by a 4-dimensional vector so that each element in this vector corresponds to the number of taxa drawn from each of the four clades $n_0,n_1,n_2,n_3$.

Next, we describe the formulas for the expected CFs under the coalescent model for every 4-taxon subset on a level-1 semi-directed network with branch lengths in coalescent units ($t_i$) and inheritance probability ($\gamma$). These formulas have already been derived in \cite{snaq}. 
Let $N$ be the network on Figure \ref{fig:4net} (top), and let $(0, 0, 2, 2)$ be the 4-taxon subset (see Figure \ref{fig:4net} (bottom right) for an example) for which we want to compute the CF formulas under the coalescent model.
Let taxa $k_1,k_2 \in n_2$ and taxa $l_1,l_2 \in n_3$.
The probability of taxa $k_1$ and $k_2$ coalescing in a branch with length $t_2+t_{23}+t_3$ is given by $1-\dfrac{2}{3} z_{2}z_{23}z_{3}$ where $z_i=\exp (-t_i)$.
Then, the formulas for the three CF under the coalescent model are defined as:
\begin{align*}
    P(k_1,k_2|l_1,l_2) &= 1-\dfrac{2}{3} z_{2}z_{23}z_{3} = a_1 \\
    P(k_1,l_1|k_2,l_2) &= \dfrac{1}{3} z_{2}z_{23}z_{3} = a_2 \\
    P(k_1,l_2|k_2,l_1) &= \dfrac{1}{3} z_{2}z_{23}z_{3} = a_3  
\end{align*}
where $a_1, a_2$ and $a_3$ denote the values for the true CFs corresponding to the splits $k_1,k_2|l_1,l_2$, $k_1,l_1|k_2,l_2$ and $k_1,l_2|k_2,l_1$ respectively. Thus, for the 4-taxon subset $(0,0,2,2)$, there are three polynomial equations that represent the probabilities for each quartet under the coalescent model. The true CF values ($a_1,a_2,a_3$) derived from these three formulas  will later be used by our algorithm as the observed CF estimated from the sample of gene trees.

In Supplementary Material, we list all of the CF formulas for all 4-taxon subsets on a level-1 semi-directed network with one 4-node hybridization cycle (Figure \ref{fig:4net} (top)). Because of the theoretical work in \cite{snaq, Solis-Lemus2020-gk}, we know that we only need two taxa per clade to represent all CF formulas involving the hybridization cycle, and thus, we can restrict to the case of $n=8$ taxa for the definition of all CF formulas (yet our method is not restricted to 8 taxa, see \nameref{sec:inf}). Furthermore, not all ${8\choose4}=70$ 4-taxon subsets are required. Indeed, we ignore 4-taxon subsets of the form $(3,1,0,0)$ since three or more taxa from one of the clades (in this case, $3$ taxa from clade $n_0$) do not provide information about the hybridization cycle. Recall that only the length of internal branches appears in the CF formulas but not the length of external branches. Therefore, if there are three or more taxa on one clade (e.g. $3$ taxa in clade $n_0$), then branches involved in the hybridization cycle will correspond to an external branch and thus, will not be included in any of the CF formulas. The same is true if all four taxa in the 4-taxon subset all come from the same clade. Thus, only 4-taxon subsets where at most $2$ taxa come from a given clade contain information about the hybridization cycle. There are $19$ such 4-taxon subsets in which $0 \leq i,j,k,l \leq 2$ for $(i,j,k,l)$, and thus, there are $19 \times 3 = 57$ concordance factor polynomial equations along with $57$ true CF values ($a_i$).

As just described, the semi-directed network $N$ in Figure \ref{fig:4net} (top) can be represented by a set of $57$ polynomial equations, denoted $CF(N)$ in $66$ variables: $9$ variables corresponding to branch lengths and inheritance probability ($z_0, z_1, z_2, z_3, z_{01}, z_{02}, z_{13}, z_{23}, \gamma$ for $z_i=\exp (-t_i)$) and $57$ variables corresponding to the true CF values ($a_i$ for $i=1,\dots,57$). Here, we identify the relationships that the $a_i$ values need to satisfy for the polynomial system $CF(N)$ to be consistent. These relationships are denoted \emph{phylogenetic invariants}.
For example, because the three concordance factors corresponding to a given 4-taxon subset (say $(0,0,2,2)$ as described above) need to sum up to one, we have one phylogenetic invariant for these three numbers: $a_1+a_2+a_3=1$, or similarly $i_1(a_1,a_2,a_3) = a_1+a_2+a_3-1$ with $i_1(a_1,a_2,a_3) = 0$. In addition, since the two minor concordance factors are equal, we have a second phylogenetic invariant defined for these numbers: $a_2=a_3$, or similarly $i_2(a_2,a_3)=a_2-a_3$ with $i_2(a_2,a_3)=0$. It turns out that the set of all concordance factor values for all 4-taxon subsets (the 57 polynomial equations denoted $CF(N)$) define a set of phylogenetic invariants (equations only in the $a_i$ variables) that need to vanish to zero whenever the concordance factors come from the true network. In this way, we can evaluate whether the given CF values come from the true network or not without knowing ($t_i$) and ($\gamma$).

Note that not all networks will have the same number of CF polynomial equations. For example, for the 8-taxon network $N=2222$, all 19 4-taxon subsets can be extracted, and thus, all 57 CF equations are defined in $CF(N)$. For the 5-taxon network $N=1112$, the 4-taxon subset $(2,1,1,0)$ cannot be considered since this subset requires two taxa from the clade $n_0$ and the 5-taxon network only has one taxon in this clade. Therefore, the 5-taxon network $N=1112$ defines a set of fewer polynomial equations (12 to be exact) in $CF(N)$ involving only the CF values: $a_7, a_8, a_9, a_{22}, a_{23}, a_{24}, a_{28}, a_{29}, a_{30}, a_{31}, a_{32}$ and $a_{33}$.

To obtain the phylogenetic invariants defined by $N$, denoted $\mathcal{I}(N)$, we get the Gr\"{o}bner basis of $CF(N)$ on the $a_i$ variables using any elimination method in Macaulay2 \cite{Grayson}. All Macaulay2 scripts (and output files) are publicly available in \url{https://github.com/solislemuslab/PhyloDiamond.jl}.

For example, for the network $N=1112$, there are 10 phylogenetic invariants in the Gr\"{o}bner basis of $CF(N)$ on the $a_i$ variables:
\begin{enumerate}
\item $a_{32} - a_{33}$
\item $a_{31} + 2a_{33} - 1$
\item $a_{28} + a_{29} + a_{30} - 1$
\item $a_{23} - a_{24}$
\item $a_{22} + 2a_{24} - 1$
\item $a_{8} - a_{9}$
\item $a_{7} + 2a_{9} - 1$
\item $3a_{9}*a_{30} + a_{9} - a_{24} - a_{33}$
\item $a_{24}*a_{29} + 2a_{24}*a_{30} + a_{29}*a_{33} - a_{30}*a_{33} - a_{33}$
\item $3a_{9}*a_{29} - 2a_{9} + 2a_{24} - a_{33}$
\end{enumerate}

We consider the first 7 invariants as the ``trivial" invariants related to the sum-to-one property and the equality of the minor CFs. The last three invariants, on the other hand, identify relationships that the true CFs need to satisfy if they originated from the $N=1112$ network. We note that the phylogenetic invariants depend on CF values $a_i$ but neither on branch lengths $t_i$ nor on inheritance probabilities $\gamma$. This property is desirable because when inferring a network, CFs values $a_i$ can be estimated from gene trees or genetic sequences, but both $t$ and $\gamma$ are unknown parameters. All the phylogenetic invariants corresponding to each $n-$taxon network ($5 \leq n \leq 8$) can be found in the Supplementary Material.

\subsection{Inference of n-taxon phylogenetic networks with one 4-node hybridization cycle using phylogenetic invariants}
\label{sec:inf}

\begin{figure*}[t!]
\centering
\includegraphics[width=1\linewidth]{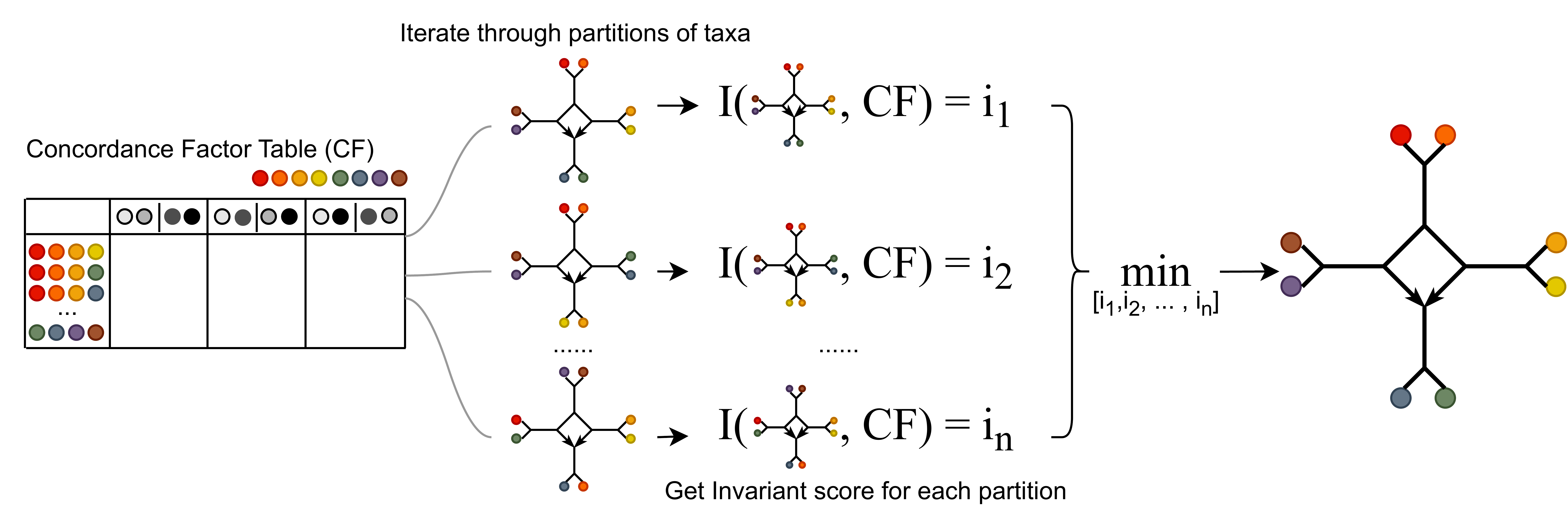}
\caption{Graphical abstract of the methodology using phylogenetic invariants to identify the best placement of the 4-node hybridization cycle to fit the table of observed concordance factors. In this case, there are 8 taxa shown as colored circles. The invariant score is defined as the $L_2$ norm of the vector of evaluated invariants.}
\label{graph-abs}
\end{figure*}

The procedure to infer a phylogenetic network with one 4-node hybridization cycle using phylogenetic invariants starts with a table of estimated CFs. As shown on Figure \ref{graph-abs} (left), the CF table contains rows representing different 4-taxon subsets and three columns labeled by grey-scale circles denoting three types of bipartitions. Each value in this table is mapped to the vector of 
$\begin{bmatrix}
    a_1 & \dots & a_{57} \\
  \end{bmatrix}$
and this vector is plugged in the invariants for each candidate network $\mathcal{I}(N)$. Here, all candidate networks are all possible partitions of given taxa. For example, as mentioned, the candidate network $N=1112$ has 10 phylogenetic invariants $(|\mathcal{I}(N)| = 10)$, and thus, after taking in the observed CFs, the invariants output a $10$-dimensional vector. The \emph{invariant score} is defined as the $L_2$ norm of the vector of evaluated invariants and the candidate network with the smallest invariant score is identified as the network that agrees best with the observed CFs. See Figure \ref{graph-abs} for a graphical abstract of the procedure and Algorithm \ref{algo8} for a more detailed description of the steps.

\begin{algorithm}[H]
\caption{Inference of a 4-node hybridization cycle in an $n$-taxon phylogenetic network ($5 \le n \le 8$) with phylogenetic invariants}
\label{algo8}
\Input{Table of estimated concordance factors; optional: the number of optimal networks ($m$) to return (default $m=5$)}
\Output{Top $m$ optimal networks with smallest invariant score}
scores $\gets$ an empty array\;

nets $\gets$ an empty array\;

\For{$N_i$, a possible partition of taxa}{
    append(nets,$N_i$)\;
    
    map observed CF values to $a_i$ values\;
    
    score $\gets$ L2norm($\mathcal{I}(N_i)$)\;
    
    append(scores, score)\;
}
rank(scores, nets)\;

\textbf{return} scores[1:m], nets[1:m]\;

\end{algorithm}

Given that we only have phylogenetic invariants for the cases of $5 \leq n \leq 8$ taxa, if the true network has $5 \leq n \leq 8$ taxa, the procedure is just as described in Figure \ref{graph-abs} and Algorithm \ref{algo8}. If the true network has more than 8 taxa, then the procedure is slightly different. In this case, our method needs to iterate through all possible subsets of 8 taxa, and fit the phylogenetic invariants on each subset to identify the top subnetwork with $n=8$ taxa. Let $N^*_8$ be the best 8-taxon network to fit the observed CFs. The missing taxa are added to $N^*_8$ according their placement in the second best network, or if the missing taxa are not in the second best network, the algorithm adds them according to their placement in the third best or the fourth best and so on. For example, assume that we have $n=9$ taxa (A,B,C,D,E,F,G,H,I) and the best network with $n=8$ taxa ($N^*_8$) contains A,B,C,D,E,F,G,H in the following partition: A,B in $n_0$; C,D in $n_1$; E,F in $n_2$, and G,H in $n_3$. We then find the best network containing taxon I. Assume that this network puts taxon I in the clade $n_2$, then the resulting best network would have E,F, and I in $n_2$. Note that our algorithm is unable to resolve the bifurcation structure within a given clade (e.g. the $n_2$ clade here with E,F, and I), but it is able to 1) correctly identify the partition of taxa among the four clades defined by the 4-node hybridization cycle and 2) correctly identify those are in the hybrid clade ($n_0$), those are in the sister clades to the hybrid clade ($n_1$ and $n_2$), and those are in the remaining fourth clade ($n_3$). See Algorithm in the Appendix for the steps in the procedure for $n>8$ taxa.

\subsection{Simulation study}

\subsubsection{Networks with eight or fewer taxa}

In this section, we focus on the case of six taxa ($N=2211, 2121, 2112, 1122, 1212, 1221$), seven taxa ($N=2221, 2212, 2122, 1222$) and eight taxa ($N=2222$). Next subsection describes the simulations with more than eight taxa.

For a given network $N$, we generate data under four settings: 1) true concordance factors (proof of concept); 2) true concordance factors perturbed by Gaussian error; 3) simulated gene trees, and 4) estimated gene trees.
To generate the true concordance factors on a given network in the first scenario, we use the Julia package \texttt{PhyloNetworks.jl} \cite{Solis-Lemus2017-jk}, and this scenario is designed to demonstrate that the invariants method works under ideal conditions. 
For the second setting, we add Gaussian noise to the true concordance factors with zero mean and standard deviation of $\sigma = 0.0005, 0.00005, 0.000005$. We choose Gaussian noise as is standard practice. 
In the third setting, we use \texttt{ms-converter} that runs \texttt{ms} \cite{ms} to simulate $k=100, 1000, 10000$ gene trees under the coalescent model on a given network $N$. All internal branches in the network are set to 1.0 coalescent unit so that the amount of incomplete lineage sorting is controlled, but not trivial, and we set an inheritance probability parameter of $\gamma=0.3$. 
For the fourth setting, we simulate sequences of length $L=500,2000$ bp on each simulated gene tree with \texttt{seq-gen} \cite{seqgen} under the HKY model, scale the branch lengths by $0.036$, and set nucleotide frequencies as $0.300414$, $0.191363$, $0.196748$, and $0.311475$. Then, we estimate gene trees with \texttt{IQ-Tree} \cite{iqtree} with ModelFinder Plus for model selection \cite{modelfinder}. We repeat each simulation setting $30$ times. All simulation scripts are in the GitHub repository \url{https://github.com/solislemuslab/PhyloDiamond.jl}.

\subsubsection{Networks with more than 8 taxa}
\label{more8}

We also analyze cases of nine taxa ($N=2223, 2232, 2322, 3222$) and ten taxa ($N=3322, 3232, 3223, 2233, 2323, 2332$). Given that the algorithm to infer a network with more than 8 taxa relies entirely on the algorithm on 8 taxa, we focus on extensive simulations for the algorithm on 8 taxa in the previous section, and simply show proof-of-concept simulations for the case of more than 8 taxa in this section. In particular, we focus on testing the following scenarios: 1) true concordance factors, and 2) true concordance factors perturbed by Gaussian error as described above ($\sigma=0.0005$).

\subsection{Reticulate evolution in the Genus \emph{Canis}}

We further test our method on the \emph{Canis} dataset from \cite{dog}. The original genomic dataset contained $12$ gray wolves, $14$ dogs, five coyotes, one Ethiopian wolf, three golden jackals, six African golden wolves, two dholes, four African hunting dogs, and one Andean fox. Given the widespread gene flow reported in the original study \cite{dog}, we need to subsample the taxa so that there is one 4-node hybridization event among the selected taxa: one African hunting dog, one coyote, one dhole, one dog, one golden jackal, and one grey wolf.
We use the estimated gene trees from \cite{astral} available in \url{https://github.com/chaoszhang/Weighted-ASTRAL_data} to infer the concordance factors with the Julia package \texttt{PhyloNetworks} \cite{Solis-Lemus2017-jk}.
We note that it is not necessary to subsample the taxa to use our method. The algorithm described for more than 8 taxa is capable of identifying the 4-node hybridization cycle for any number of taxa. However, as mentioned, its limitation is that we are unable to resolved the relationships within the four clades. In this case, we decided to subsample the species in order to obtain a fully resolved phylogeny.

To better evaluate our method's performance, we also run other methods on this dataset for comparison. In addition to our method, we also run \texttt{SNaQ} \cite{snaq} on $h=1$ hybridization event, 10 independent runs, and using one randomly selected gene tree as the starting tree. We use all 449,450 gene trees to infer the table of CFs and then map the allele names to species names. We also run \texttt{PhyloNet} on two options: 1) maximum likelihood (\texttt{ML}) and 2) maximum pseudolikelihood (\texttt{MPL}). We choose $h=1$ hybridization event on both cases, with 10 independent runs for each. We choose one representative individual per species in the gene trees and removed gene trees that contained fewer than 5 taxa. In total, we use 448,758 gene trees in the \texttt{PhyloNet} analyses. We root the gene trees on the known outgroup (African hunting dog) or on dhole if African hunting dog is not in the gene tree.



\section{Results}
\label{results}

\subsection{Simulation study on eight or fewer taxa}

\subsubsection{Identifying the correct network as the top 1 ranked network with the smallest invariant score}

\begin{figure}[h!]
\centering
\includegraphics[scale=0.2]{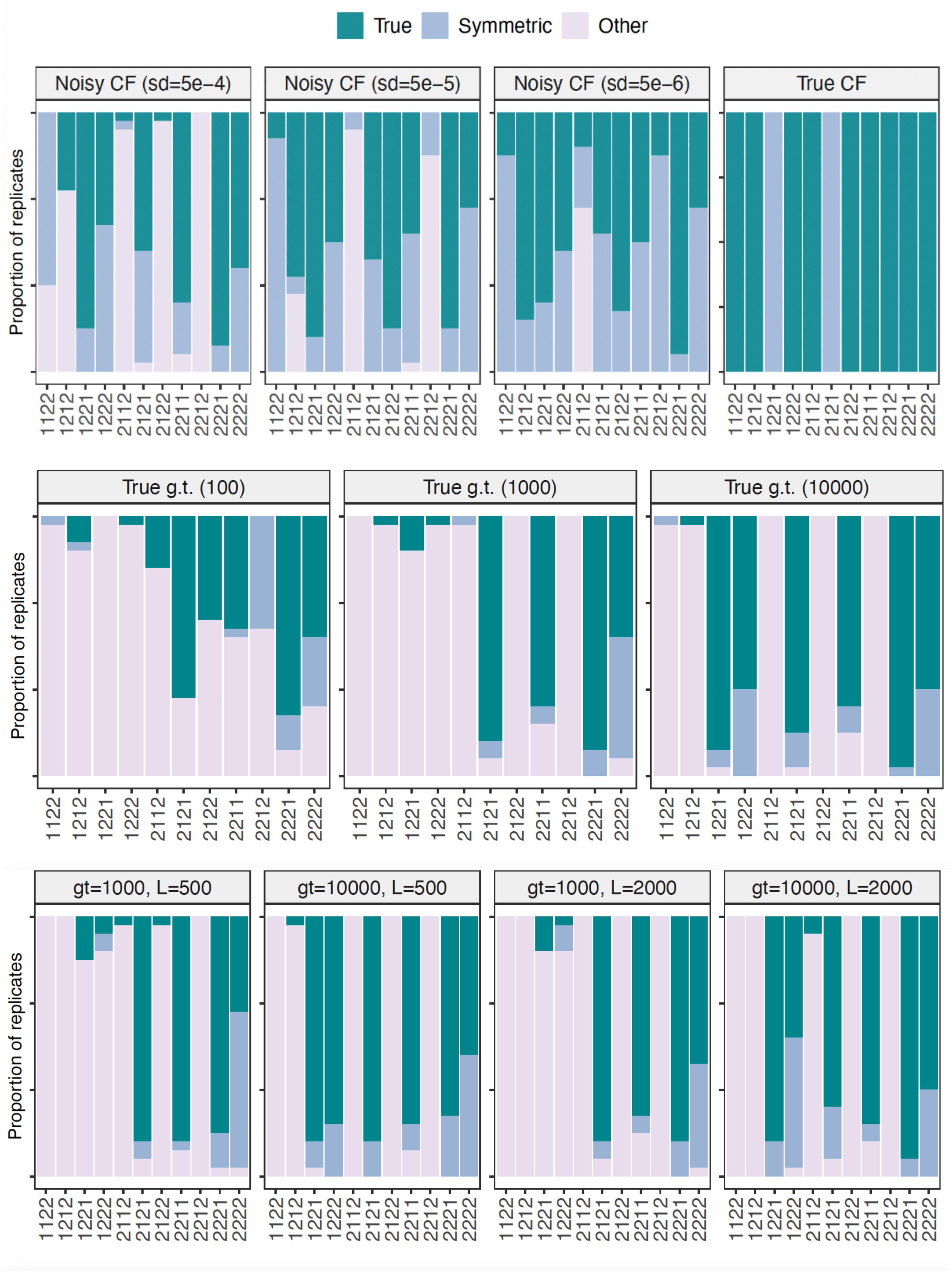}
\caption{Proportion of times that the true hybridization cycle or the symmetric cycle (with clades $n_1$ and $n_2$ inverted) are in the top 1 ranked hybridization cycles identified by the phylogenetic invariants on 3 different cases. Y-axis is between 0 and 100. \textbf{First row}: true and Gaussian-perturbed CFs. Each panel corresponds to a type of simulation: using true concordance factors (left) and using concordance factors with added Gaussian noise (with increasing standard deviation for noise from left to right). 
Whenever there are at least two taxa in each of the four clades (network $2222$), our method is very accurate in detecting the true (or its symmetric) network. Our method is also accurate to identify the true network in the top 5 ranked networks (see Figure \ref{top5} (first row)) which will greatly reduce the space of candidate networks to be compared with a likelihood approach. 
\textbf{Second row}: true simulated gene trees (``g.t."). 
Whenever there are at least two taxa in each of the four clades (network $2222$), our method is very accurate in detecting the true (or its symmetric) network. Our method is also accurate to identify the true network in the top 5 ranked networks (see Figure \ref{top5} (second row)) which will greatly reduce the space of candidate networks to be compared with a likelihood approach. 
\textbf{Third row}: estimated gene trees. Each panel corresponds to a number of gene trees (g.t. from 1000 to 10,000) and sequence length ($L$ from 500bp to 2000bp). 
Whenever there are at least two taxa in each of the four clades (network $2222$) or only clade $n_3$ having one taxon ($N=2221$), our method is very accurate in detecting the true (or its symmetric) network. Our method is also accurate to identify the true network in the top 5 ranked networks (see Figure \ref{top5} (third row)) which will greatly reduce the space of candidate networks to be compared with a likelihood approach.
}
\label{top1}
\end{figure}

Figure \ref{top1} (first row) shows the proportion of times (out of the 30 replicates) that our invariants method identifies the true network as the top 1 ranked network with the smallest invariant score for the cases of true concordance factors and the Gaussian-perturbed CFs with increasing standard deviation (from left to right). We also quantify the proportion of times that our invariants method identifies the symmetric network (when clades $n_1$ and $n_2$ in Figure \ref{fig:4net} (top) are switched) because in the 4-node hybridization cycle it is difficult for the method to distinguish clades $n_1$ and $n_2$ given that they both contribute to the hybridization node.
Even for the case of true CFs, our method identifies the symmetric network as the top 1 network, instead of the true network, in two cases ($N=1221$ and $N=2121$), yet the difference in invariance score ($L_2$-norm of evaluated polynomials) of the true and symmetric networks in these cases is of the order of $10^{-16}$.
For the noisiest setting ($\sigma=0.0005$, far right), it is evident that some networks are difficult to be detected by our method, but it is worth highlighting that whenever there are two taxa sampled on each of the four clades ($n_0, n_1, n_2, n_3$) as in $N=2222$, our method always identifies the true (or symmetric) network as the top 1. This condition can easily be satisfied whenever there are multiple individuals per species.

Figure \ref{top1} (second row) shows the proportion of times (out of the 30 replicates) that our invariants method identifies the true network as the top 1 ranked network with smallest invariant score for the case of true simulated gene trees with increasing number of gene trees (from left to right), where ``g.t." is abbreviation for ``gene trees". Unlike the case of true or Gaussian-perturbed CFs, our method only identifies the true (or symmetric) network when there are 2 taxa sampled per clade ($N=2222$) or at least 2 taxa sampled on the hybrid clade and sister clades ($N=2221$). With 1,000 gene trees or more, the networks $N=2121$ and $N=2211$ are also accurately identified by our method.

Figure \ref{top1} (third row) shows the proportion of times (out of the 30 replicates) that our invariants method identifies the true network as the top 1 ranked network with smallest invariant score for the case of estimated gene trees with increasing number of gene trees and sequence length (from left to right). Again, our method only identifies the true (or symmetric) network when there are 2 taxa sampled per clade ($N=2222$) or at least 2 taxa sampled on the hybrid clade and sister clades ($N=2221$). As before, with 1,000 gene trees, the networks $N=2121$ and $N=2211$ are also accurately identified by our method, and with 10,000 gene trees, the networks $N=1221$ and $N=1222$ are also accurately identified.

Summarizing Figure \ref{top1} related to our method's ability to identify the true (or symmetric) network as top 1 ranked network by its invariant score, we can highlight that networks with only one sampled taxon in some of the clades are harder to be identified. As long as two taxa are sampled from each of the clades, our method is able to identify the network with high accuracy on all simulation settings. The number of genes is more important than the sequence length, and with at least 100 genes, our method is able to correctly identify networks with fewer than two taxa sampled from some clades (like $2221$). We highlight that we are equally interested in the true and symmetric networks since a follow-up optimization of branch lengths and inheritance probability ($\gamma$) on the fixed network will estimate the correct $\gamma$ and identify which of the sister clades is the major ($\gamma > 0.5$) and which is the minor ($\gamma <0.5$).

\subsubsection{Reducing the space of candidate networks using the invariant score}

\begin{figure}[h!]
\centering
\includegraphics[scale=0.2]{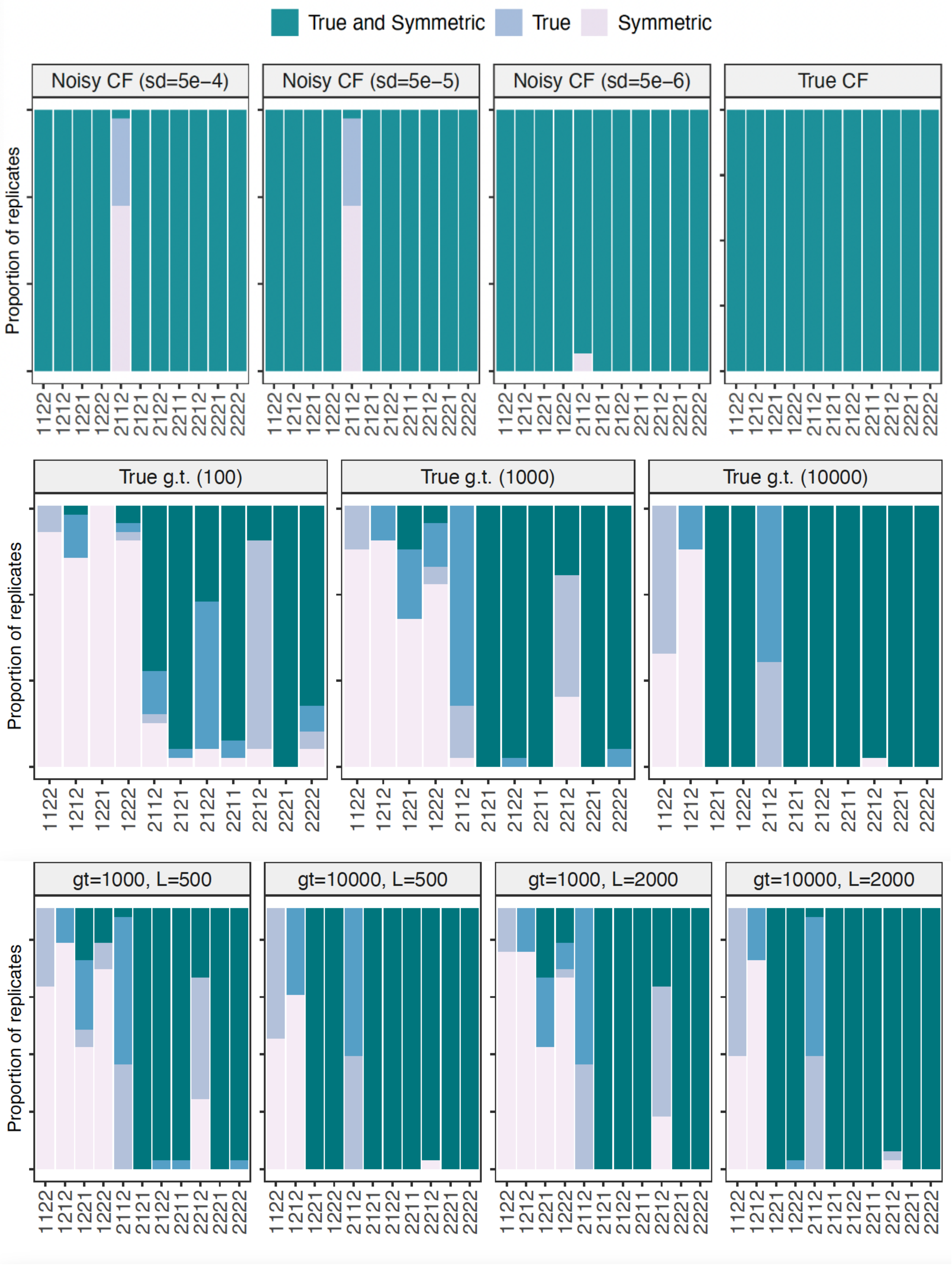}
\caption{Proportion of times that the true network or the symmetric network (with clades $n_1$ and $n_2$ inverted) are in the top 5 ranked networks identified by the phylogenetic invariants on 3 different cases. Y-axis is between 0 and 100. \textbf{First row}:true and Gaussian-perturbed CFs. Each panel corresponds to a type of simulation: using true concordance factors (left) and using concordance factors with added Gaussian noise (with increasing standard deviation for noise from left to right). 
All networks are accurately identified in the top 5 ranked based on invariant score which provides evidence that our method fast and accurately reduce the space of candidate networks to just 5 alternatives.
\textbf{Second row}: true simulated gene trees (``g.t.").
Only the networks that have one taxon below the hybrid node (1122, 1212, 1221, 1222) do not allow accurate reconstruction with fewer than 10,000 gene trees which brings into attention the importance of taxon sampling for this method.
\textbf{Third row}:estimated gene trees. Each panel corresponds to a number of gene trees (g.t. from 1000 to 10,000) and sequence length ($L$ from 500bp to 2000bp). 
Only the networks that have one taxon below the hybrid node (1122, 1212, 1221, 1222) do not allow accurate reconstruction with fewer than 10,000 gene trees which brings into attention the importance of taxon sampling for this method.
}
\label{top5}
\end{figure}

Figure \ref{top5} (first row) shows the proportion of times (out of the 30 replicates) that the true (or symmetric) network are within the top 5 ranked networks based on smallest invariant score for the cases of true concordance factors and the Gaussian-perturbed CFs with increasing standard deviation (from left to right). In all cases, our method includes the true (and symmetric) networks within the top 5 which means that our method accurately and in a fast manner reduces the space of candidate networks to just 5 candidates that can later be tested with another accurate methodology like likelihood.

Figure \ref{top5} (second row) shows the proportion of times (out of the 30 replicates) that the true (or symmetric) network are within the top 5 ranked networks based on smallest invariant score for the case of true simulated gene trees. Only the networks with one sampled taxon on the hybrid clade ($n_0$) cannot be recovered with 100 or 1000 gene trees. Whenever there are 10,000 gene trees in the sample, the networks $N=1221$ and $N=1222$ can now be recovered as part of the top 5, yet the networks $N=1122$ or $N=1212$ (when there is one sampled taxon in the hybrid clade and one of the sister clades) failed to be detected.
The same conclusions are true for the case of estimated gene trees (Figure \ref{top5} (third row)) with number of gene trees having more weight on the accuracy than sequence length.

Summarizing Figure \ref{top5} related to our method's ability to identify the true (or symmetric) network within the top 5 ranked networks by its invariant score, we can highlight that as long as there are at least 2 taxa sampled from the hybrid clade, then our method is able to place the true (and symmetric) network within the top 5 networks. This effectively reduces the space of candidate networks to compare with a likelihood approach. As mentioned, we do not view our method as a substitute to existing network inference methods. On the contrary, we believe that our invariants method will serve as a complement to identify a small subset of network possibilities that will help bypass the optimization on network space.

\begin{figure}[h!]
\centering
\includegraphics[scale=0.2]{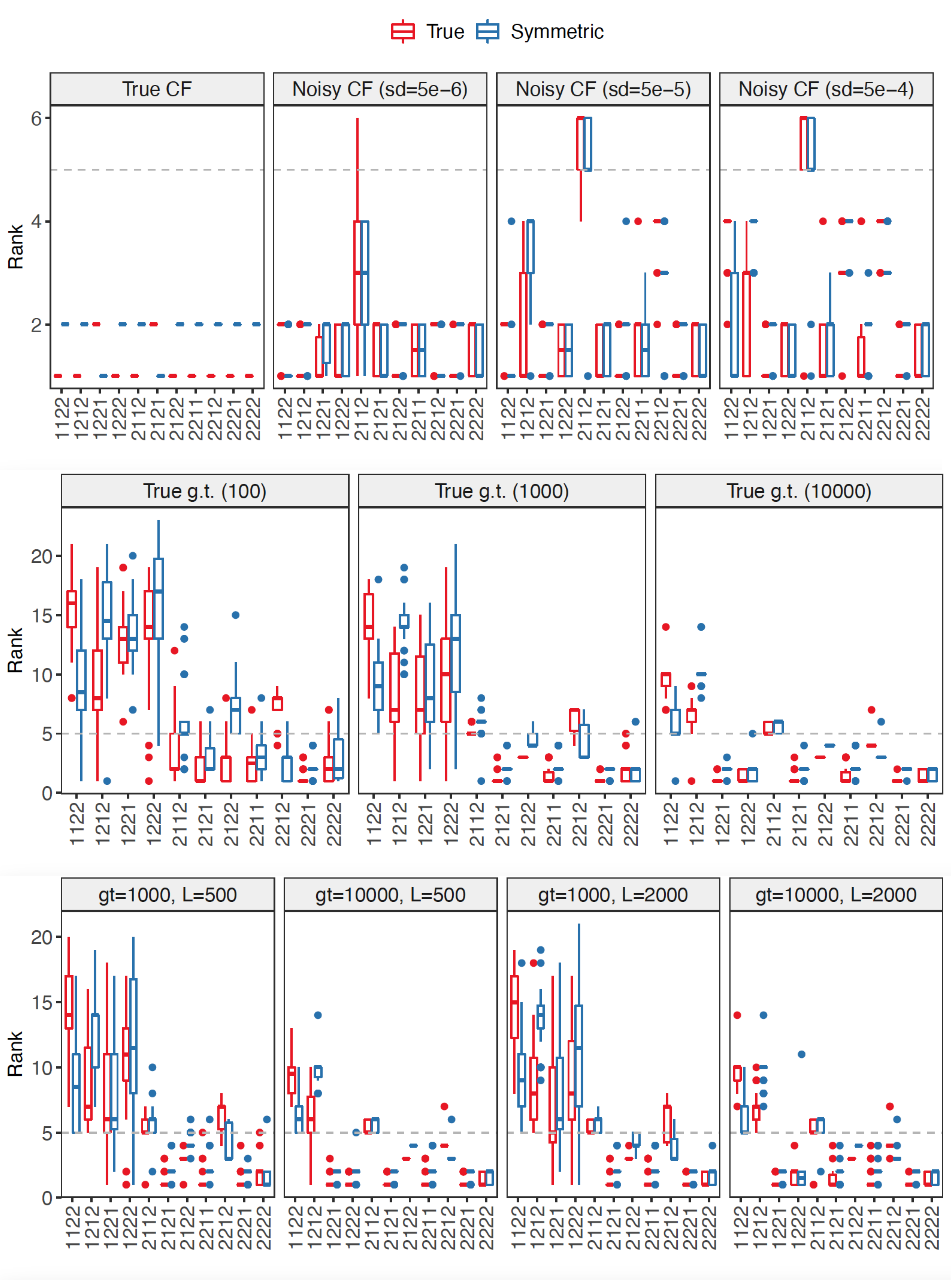}
\caption{Rank (y-axis) in the invariant score for each network (x-axis) for the true network (red) and its symmetric network (inverted clades $n_1$ and $n_2$, in blue) on 3 different cases. Dashed line corresponds to rank 5.
\textbf{First row}: true and Gaussian-perturbed CFs.  Each panel corresponds to a type of simulation: using true concordance factors (left) and using concordance factors with added Gaussian noise (with increasing standard deviation for noise from left to right). 
Both the true and symmetric networks are within the top 5 ranked networks by the method in all cases, and are thus, easy to distinguish from wrong networks.
\textbf{Second row}: true simulated gene trees (``g.t"). Each panel corresponds to a number of simulated gene trees from 100 (left) to 10,000 (right). 
Both the true and symmetric networks are within the top 5 ranked networks by the method as the number of genes increases, and are thus, easy to distinguish from wrong networks. Only the networks that have one taxon below the hybrid node (1122, 1212, 1221, 1222) do not allow accurate reconstruction which brings into attention the importance of taxon sampling for this method.
\textbf{Third row}:estimated gene trees. Each panel corresponds to a number of gene trees (g.t. from 1000 to 10,000) and sequence length ($L$ from 500bp to 2000bp). Only the networks that have one taxon below the hybrid node (1122, 1212, 1221, 1222) do not allow accurate reconstruction which brings into attention the importance of taxon sampling for this method.
}
\label{rank}
\end{figure}

Figure \ref{rank} (first row) shows the rank in the invariant scores of the true and symmetric networks on the cases of true and Gaussian-perturbed CFs. Ideally, the true network (or at least its symmetric version) should be ranked as 1 by our invariants method. While the true and symmetric networks are not always ranked in number 1, they are ranked within the top 5 for most of the cases. This further confirms the advantage of our current method. It is a fast way to reduce the space of candidate networks (to 5) that can later be tested using a likelihood approach.

Figure \ref{rank} (second row) shows the rank in the invariant scores of the true and symmetric networks for the case of true simulated gene trees (increased number of gene trees from left to right). The true (and symmetric) networks are not within the top 5 networks only for the cases of one sampled taxon from the hybrid clade ($N=1122,1212,1221,1222$). This same behavior is evident in Figure \ref{rank} (third row) for estimated gene trees. Again, these figures show our method's ability to reduce the possible networks that fit the data.

In Supplementary Material, we also show plots with the values of the invariant scores for the true and symmetric networks.

Last, we present a comparison of the running times of four network methods in the Appendix. While all methods are able to identify the correct network, our phylogenetic invariants method (top row) only takes 7.17 seconds to correctly infer the network $N=2222$ which is over twice as fast as the second fastest (\texttt{PhyloNet MPL} \cite{yu2015maximum}). The difference in running times (and even accuracy) is more evident in the results on the \emph{Canis} dataset (Section \ref{dog-results}).

\subsection{Simulation study on more than eight taxa}

Table in the Supplementary Material shows the proof-of-concept results on simulated data with more than 8 taxa. Since the algorithm for more than 8 taxa (Algorithm in the Supplementary Material) relies entirely on the algorithm for 8 or fewer taxa (Algorithm \ref{algo8}), we simply show here that our steps to append taxon to the optimal subset of 8 taxa identified by the method indeed yields the desired network with more than 8 taxa.

\subsection{Phylogenetic network for the \textit{Canis} genus}
\label{dog-results}

\begin{figure*}[hbt!]
\centering
\includegraphics[scale=0.2]{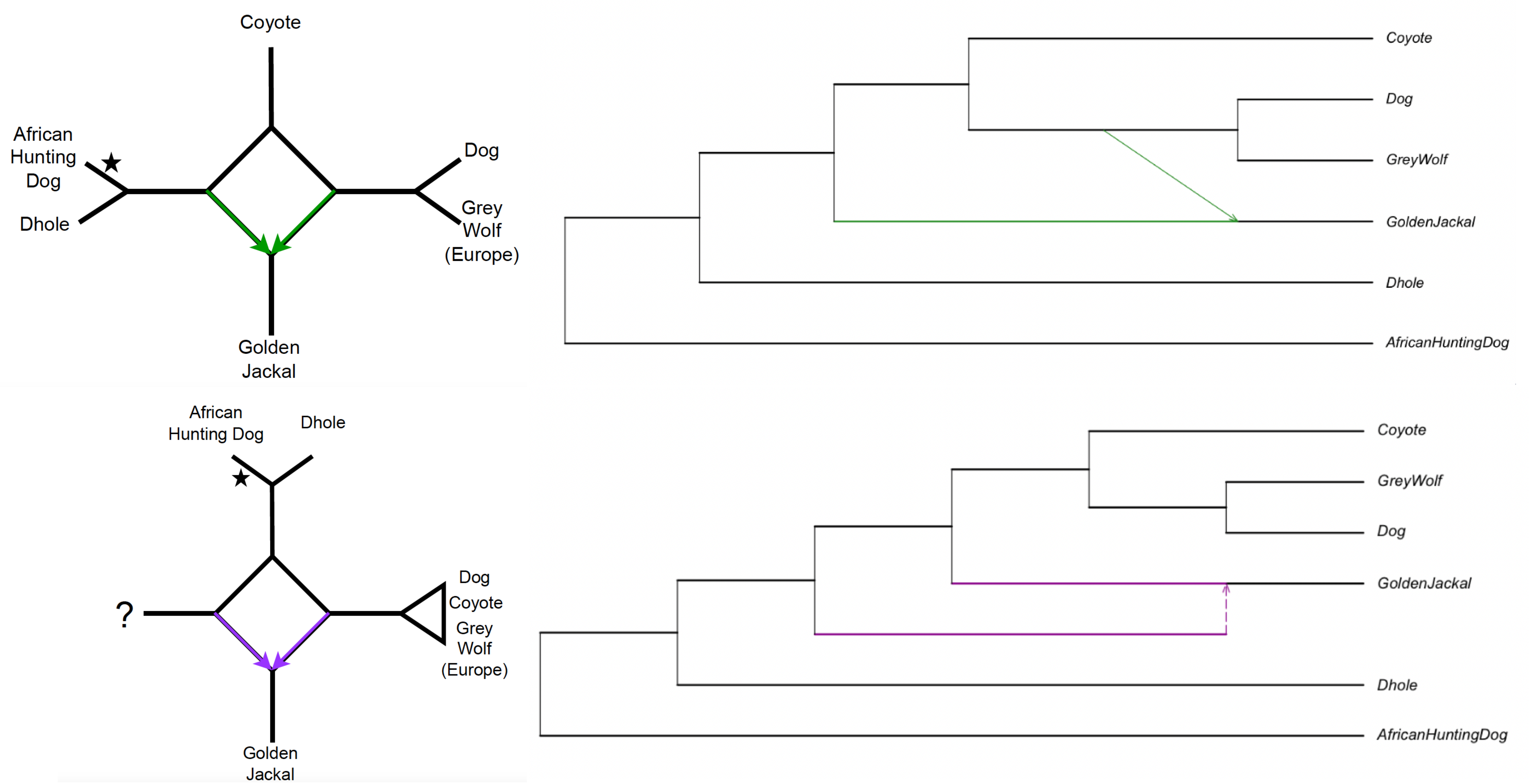}
\caption{\textbf{Left top}: Semi-directed phylogenetic network estimated by \texttt{SNaQ} \cite{snaq}, \texttt{PhyloNet ML} \cite{Yu2014}, and our method based on phylogenetic invariants. 
\textbf{Left bottom}: Semi-directed phylogenetic network estimated by \texttt{PhyloNet MPL} \cite{yu2015maximum}. The star marks the place where the root should be. 
\textbf{Right}: Rooted version of the semi-directed networks. 
The network estimated with \texttt{PhyloNet MPL} (bottom) is different from the network inferred by the other methods, and it involves an unsampled (or extinct) taxon in the hybridization. \texttt{SNaQ}, \texttt{PhyloNet ML} and our phylogenetic invariants method (top) identify the same hybridization event as the original publication \cite{dog} between the ancestral species to dog and grey wolf and the ancestral species of golden jackal, yet \texttt{PhyloNet ML} takes 389 times longer and \texttt{SNaQ} takes 20 times longer than our invariants method (see running times in Table \ref{dog-time}).}
\label{dog-net}
\end{figure*}

The original analysis of the \emph{Canis} genus identified nine gene flow events using the D statistics \cite{pattersonD} (Figure 3a in \cite{dog}). Here, we are able to replicate the hybridization event involving gene flow from the ancestor of dog and grey wolf into the ancestor of golden jackal (Figure \ref{dog-net} (top)). The same network is correctly inferred by \texttt{SNaQ} and by \texttt{PhyloNet ML}, but both methods take much longer to run (Table \ref{dog-time}). \texttt{PhyloNet MPL} identifies a different hybridization event that has not been reported in previous studies, so it is impossible to validate it at this point. This hybridization involves an unsampled or extinct taxon (Figure \ref{dog-net} (bottom)).

In terms of running time, our invariants method is able to identify the correct hybridization in under 7 seconds while \texttt{SNaQ} (which also identifies the correct hybridization) takes over 140 seconds. \texttt{PhyloNet MPL} takes 40 times longer than our invariants method (with running time of over 281 seconds) and identifies a different hybridization event to those originally published (so, not possible to validate at this point). \texttt{PhyloNet ML} is the slowest method with over 45 minutes of running time on this small network of only 6 taxa. The inferred network by \texttt{PhyloNet ML} also agrees with the one originally published in \cite{dog}.

\begin{table}[h!]
  \begin{center}
    \begin{tabular}{l|c}
      \toprule 
      \textbf{Method} & \textbf{Time (seconds)} \\
      Phylogenetic invariants (our method)    & 6.78 \\
      \texttt{SNaQ}          & 140.58 \\
      \texttt{PhyloNet ML}   & 2723.99 \\
      \texttt{PhyloNet MPL}  & 281.25 \\
      \bottomrule 
    \end{tabular}
    \caption{Running times (in seconds) on the \textit{Canis} dataset. Our method based on evaluation of phylogenetic invariants is 20 times faster than the second fastest method (\texttt{SNaQ}).}
    \label{dog-time}
  \end{center}
\end{table}

\section{Discussion}
\label{sec:disc}

Here, we introduce a novel method to infer 4-node hybridization cycles on level-1 phylogenetic networks of over five taxa based on ultrafast evaluations of phylogenetic invariants. Our method bypasses optimization on network space and is able to accurately detect the hybridization cycles based on simulated and real data, especially when there are at least 2 taxa sampled from each of the four clades defined by the hybridization cycle (Figure \ref{fig:4net} (top)). 
While our method is at least $\sim 10$ times faster than other existing network methods, it is not meant to replace existing methods. On the contrary, we believe that our ultrafast algorithm can work in conjunction with existing methodologies by reducing the space of candidate networks that can later be evaluated based on likelihood or Bayesian approaches.

Nevertheless, there are limitations in our current method. Specifically, we are only able to detect hybridization cycles with 4 nodes on level-1 networks. For the case of three nodes in the hybridization cycle, it has been investigated already in \cite{snaq} that the set of phylogenetic invariants is empty. However, there is still room to extend the current methodology to hybridization cycles of five nodes or more. 
In addition, even when our method can infer networks with any number of taxa, it is unable to resolve the topology inside any of the four clades defined by the hybridization cycle (Figure \ref{fig:4net} (top)). That is, our method is only able to identify the placement of the 4-node hybridization cycle. For this reason, we view our method as in line with other hybrid detection methods such as ABBA-BABA test \cite{pattersonD}, MSCquartets \cite{Allman-Singularities-and-Boundaries, MSCquartets-RPackage}, and HyDe \cite{hyde-paper} that are also only able to detect specific hybridization patterns in subsets of taxa.

Future work will involve the generation of the phylogenetic invariants related to hybridization cycles of five nodes or more, and the development of a merging algorithm that can produced a fully resolved $n$-taxon phylogenetic network from the 8-taxon estimated networks inferred by our invariants method.
In addition, our proposed method is unable to infer biological parameters like inheritance probability or branch length, but future work could exploit the CF formulas (which depend on branch lengths and inheritance probabilities) to provide estimated values to these parameters from the observed CFs.

\section{Data and code availability}
The \emph{Canis} dataset was made publicly available by the original publication \cite{dog} and can be accessed through the GitHub repository of \cite{astral} here \url{https://github.com/chaoszhang/Weighted-ASTRAL_data}. All the scripts for our work are publicly available in the GitHub repository \url{https://github.com/solislemuslab/PhyloDiamond.jl}. The new Julia package \texttt{PhyloDiamond.jl} is open-source, publicly available in \url{https://github.com/solislemuslab/PhyloDiamond.jl}.

\section{Acknowledgements}
This work was supported by the National Science Foundation [DEB-2144367 to CSL]. This was also funded by the UW-Madison Fall Competition [to CSL].
The authors thank Marianne Bj{\o}rner and Nathan Kolbow for help with phylogenetics code, and Sam Ozminkowski for help with building the Julia package.

\bibliographystyle{plain}
\bibliography{references}  

\newpage
\appendix

\section{Representation of a level-1 semi-directed phylogenetic network with a set of polynomial equations corresponding to the expected concordance factors under the coalescent model}
\label{expCF}

Notation: 
\begin{itemize}
    \item $n$ corresponds to the number of individuals from each of the 4 clades: $(n_0, n_1, n_2, n_3)$. For example, $n=(0,0,2,2)$ means that you have the quartet with 2 individuals in $n_2$ and 2 individuals in $n_3$
    \item "Individuals" list the individuals taken from each of the four clades. For example, $i_1,i_2 \in n_0$ means that we took two species from clade $n_0$ denoted $i_1$ and $i_2$. We use these species names to define the splits for the formulas: e.g. $P(i_1,i_2|j_1,j_2)$ represents the probability that $i_1$ and $i_2$ are together in one side of the split (and $j_1,j_2$ together)
    \item "Type" corresponds to the type of quartet which matches the types in \cite{snaq}
    \item "CF Formula" corresponds to the formula of the expected CF for that given quartet under the multispecies coalescent model
    \item "CF value" is the variable we give to the observed CF we will read from the data table
\end{itemize}

By \cite{Solis-Lemus2020-gk}, we know that we only need to select at most two individuals per subgraph $(n_0, n_1, n_2, n_3)$ to define all the CF formulas that involve the hybridization cycle. We list all the CF formulas in the table below assuming that we do have 2 individuals per clade $(n_0, n_1, n_2, n_3)$ (that is, we have at least eight species).

\begin{longtable}{@{\extracolsep{\fill}}|l|l|l|l|@{}}
	\hline
		$ n $ & Type & CF Formula  & CF value \\ \hline
		$ (0,0,2,2) $ & $ 5 $ & $ 1-\dfrac{2}{3} z_{2}z_{2,3}z_{3} $ & $ a1 $ \\ 
		& & $ \dfrac{1}{3} z_{2}z_{2,3}z_{3} $ & $ a2 $ \\ 
		& & $ \dfrac{1}{3} z_{2}z_{2,3}z_{3} $ & $ a3 $ \\ \hline
		$ (0,1,2,1) $ & $ 5 $ & $ 1-\dfrac{2}{3} z_{2,3}z_{2} $ & $ a4 $ \\ 
		& & $ \dfrac{1}{3} z_{2,3}z_{2} $ & $ a5 $ \\ 
		& & $ \dfrac{1}{3} z_{2,3}z_{2} $ & $ a6 $ \\ \hline
		$ (0,1,1,2) $ & $ 5 $ & $ 1-\dfrac{2}{3} z_{3} $ & $ a7 $ \\ 
		& & $ \dfrac{1}{3} z_{3} $ & $ a8 $ \\ 
		& & $ \dfrac{1}{3} z_{3} $ & $ a9 $ \\ \hline
		$ (0,2,2,0) $ & $ 5 $ & $ 1-\dfrac{2}{3} z_{2}z_{2,3}z_{1,3}z_{1} $ & $ a10 $ \\ 
		& & $ \dfrac{1}{3} z_{2}z_{2,3}z_{1,3}z_{1} $ & $ a11 $ \\ 
		& & $ \dfrac{1}{3} z_{2}z_{2,3}z_{1,3}z_{1} $ & $ a12 $ \\ \hline
  		$ (0,2,1,1) $ & $ 5 $ & $ 1-\dfrac{2}{3} z_{1,3}z_{1} $ & $ a13 $ \\ 
		& & $ \dfrac{1}{3} z_{1,3}z_{1} $ & $ a14 $ \\ 
		& & $ \dfrac{1}{3} z_{1,3}z_{1} $ & $ a15 $ \\ \hline
		$ (0,2,0,2) $ & $ 5 $ & $ 1-\dfrac{2}{3} z_{3}z_{1,3}z_{1} $ & $ a16 $ \\ 
		& & $ \dfrac{1}{3} z_{3}z_{1,3}z_{1} $ & $ a17 $ \\ 
		& & $ \dfrac{1}{3} z_{3}z_{1,3}z_{1} $ & $ a18 $ \\ \hline
		$ (1,0,2,1) $ & $ 2 $ & $ (1- \gamma )\left(1-\dfrac{2}{3} z_{2,3}z_{2} \right) + \gamma \left(1-\dfrac{2}{3} z_{2} \right) $ & $ a19 $ \\ 
		& & $ (1- \gamma )\dfrac{1}{3} z_{2,3}z_{2} + \gamma \dfrac{1}{3} z_{2} $ & $ a20 $ \\ 
		& & $ (1- \gamma )\dfrac{1}{3} z_{2,3}z_{2} + \gamma \dfrac{1}{3} z_{2} $ & $ a21 $ \\ \hline
		$ (1,0,1,2) $ & $ 2 $ & $ (1- \gamma )\left(1-\dfrac{2}{3} z_{3} \right) + \gamma \left(1-\dfrac{2}{3} z_{2,3}z_{3} \right) $ & $ a22 $ \\ 
		& & $ (1- \gamma )\dfrac{1}{3} z_{3} + \gamma \dfrac{1}{3} z_{2,3}z_{3} $ & $ a23 $ \\ 
		& & $ (1- \gamma )\dfrac{1}{3} z_{3} + \gamma \dfrac{1}{3} z_{2,3}z_{3} $ & $ a24 $ \\ \hline
		$ (1,1,2,0) $ & $ 2 $ & $ (1- \gamma )\left(1-\dfrac{2}{3} z_{1,3}z_{2,3}z_{2} \right) + \gamma \left(1-\dfrac{2}{3} z_{2} \right) $ & $ a25 $ \\ 
		& & $ (1- \gamma )\dfrac{1}{3} z_{1,3}z_{2,3}z_{2} + \gamma \dfrac{1}{3} z_{2} $ & $ a26 $ \\ 
		& & $ (1- \gamma )\dfrac{1}{3} z_{1,3}z_{2,3}z_{2} + \gamma \dfrac{1}{3} z_{2} $ & $ a27 $ \\ \hline
		$ (1,1,1,1) $ & $ 3 $ & $ (1- \gamma )\left(1-\dfrac{2}{3} z_{1,3} \right) + \gamma \dfrac{1}{3} z_{2,3} $ & $ a28 $ \\ 
		& & $ (1- \gamma )\dfrac{1}{3} z_{1,3} + \gamma \left(1-\dfrac{2}{3} z_{2,3} \right) $ & $ a29 $ \\ 
		& & $ (1- \gamma )\dfrac{1}{3} z_{1,3} + \gamma \dfrac{1}{3} z_{2,3} $ & $ a30 $ \\ \hline
		$ (1,1,0,2) $ & $ 2 $ & $ (1- \gamma )\left(1-\dfrac{2}{3} z_{1,3}z_{3} \right) + \gamma \left(1-\dfrac{2}{3} z_{3} \right) $ & $ a31 $ \\ 
		& & $ (1- \gamma )\dfrac{1}{3} z_{1,3}z_{3} + \gamma \dfrac{1}{3} z_{3} $ & $ a32 $ \\ 
		& & $ (1- \gamma )\dfrac{1}{3} z_{1,3}z_{3} + \gamma \dfrac{1}{3} z_{3} $ & $ a33 $ \\ \hline
		$ (1,2,1,0) $ & $ 2 $ & $ (1- \gamma )\left(1-\dfrac{2}{3} z_{1} \right) + \gamma \left(1-\dfrac{2}{3} z_{2,3}z_{1,3}z_{1} \right) $ & $ a34 $ \\ 
		& & $ (1- \gamma )\dfrac{1}{3} z_{1} + \gamma \dfrac{1}{3} z_{2,3}z_{1,3}z_{1} $ & $ a35 $ \\ 
		& & $ (1- \gamma )\dfrac{1}{3} z_{1} + \gamma \dfrac{1}{3} z_{2,3}z_{1,3}z_{1} $ & $ a36 $ \\ \hline
		$ (1,2,0,1) $ & $ 2 $ & $ (1- \gamma )\left(1-\dfrac{2}{3} z_{1} \right) + \gamma \left(1-\dfrac{2}{3} z_{1,3}z_{1} \right) $ & $ a37 $ \\ 
		& & $ (1- \gamma )\dfrac{1}{3} z_{1} + \gamma \dfrac{1}{3} z_{1,3}z_{1} $ & $ a38 $ \\ 
		& & $ (1- \gamma )\dfrac{1}{3} z_{1} + \gamma \dfrac{1}{3} z_{1,3}z_{1} $ & $ a39 $ \\ \hline
		$ (2,0,2,0) $ & $ 4 $ & $ (1- \gamma )^2\left(1-\dfrac{2}{3} z_{2}z_{0}z_{0,1}z_{1,3}z_{2,3} \right) +2 \gamma (1- \gamma )\left(1-\dfrac{2}{3} z_{2}z_{0} \right) + \gamma ^2\left(1-\dfrac{2}{3} z_{2}z_{0}z_{0,2} \right) $ & $ a40 $ \\ 
		& & $ (1- \gamma )^2\dfrac{1}{3} z_{2}z_{0}z_{0,1}z_{1,3}z_{2,3} +2 \gamma (1- \gamma )\dfrac{1}{3} z_{2}z_{0} + \gamma ^2\dfrac{1}{3} z_{2}z_{0}z_{0,2} $ & $ a41 $ \\ 
		& & $ (1- \gamma )^2\dfrac{1}{3} z_{2}z_{0}z_{0,1}z_{1,3}z_{2,3} +2 \gamma (1- \gamma )\dfrac{1}{3} z_{2}z_{0} + \gamma ^2\dfrac{1}{3} z_{2}z_{0}z_{0,2} $ & $ a42 $ \\ \hline
		$ (2,0,1,1) $ & $ 1 $ & $ (1- \gamma )^2\left(1-\dfrac{2}{3} z_{0}z_{1,3}z_{0,1} \right) +2 \gamma (1- \gamma )\left(1- z_{0} +\dfrac{1}{3} z_{0}z_{2,3} \right) + \gamma ^2\left(1-\dfrac{2}{3} z_{0}z_{0,2} \right) $ & $ a43 $ \\ 
		& & $ (1- \gamma )^2\dfrac{1}{3} z_{0}z_{1,3}z_{0,1} + \gamma (1- \gamma ) z_{0} \left(1-\dfrac{1}{3} z_{2,3} \right) + \gamma ^2\dfrac{1}{3} z_{0}z_{0,2} $ & $ a44 $ \\ 
		& & $ (1- \gamma )^2\dfrac{1}{3} z_{0}z_{1,3}z_{0,1} + \gamma (1- \gamma ) z_{0} \left(1-\dfrac{1}{3} z_{2,3} \right) + \gamma ^2\dfrac{1}{3} z_{0}z_{0,2} $ & $ a45 $ \\ \hline
		$ (2,0,0,2) $ & $ 4 $ & $ (1- \gamma )^2\left(1-\dfrac{2}{3} z_{3}z_{0}z_{1,3}z_{0,1} \right) +2 \gamma (1- \gamma )\left(1-\dfrac{2}{3} z_{3}z_{0} \right) + \gamma ^2\left(1-\dfrac{2}{3} z_{3}z_{0}z_{2,3}z_{0,2} \right) $ & $ a46 $ \\ 
		& & $ (1- \gamma )^2\dfrac{1}{3} z_{3}z_{0}z_{1,3}z_{0,1} +2 \gamma (1- \gamma )\dfrac{1}{3} z_{3}z_{0} + \gamma ^2\dfrac{1}{3} z_{3}z_{0}z_{2,3}z_{0,2} $ & $ a47 $ \\ 
		& & $ (1- \gamma )^2\dfrac{1}{3} z_{3}z_{0}z_{1,3}z_{0,1} +2 \gamma (1- \gamma )\dfrac{1}{3} z_{3}z_{0} + \gamma ^2\dfrac{1}{3} z_{3}z_{0}z_{2,3}z_{0,2} $ & $ a48 $ \\ \hline
		$ (2,1,1,0) $ & $ 1 $ & $ (1- \gamma )^2\left(1-\dfrac{2}{3} z_{0}z_{0,1} \right) +2 \gamma (1- \gamma )\left(1- z_{0} +\dfrac{1}{3} z_{0}z_{2,3}z_{1,3} \right) + \gamma ^2\left(1-\dfrac{2}{3} z_{0}z_{0,2} \right) $ & $ a49 $ \\ 
		& & $ (1- \gamma )^2\dfrac{1}{3} z_{0}z_{0,1} + \gamma (1- \gamma ) z_{0} \left(1-\dfrac{1}{3} z_{2,3}z_{1,3} \right) + \gamma ^2\dfrac{1}{3} z_{0}z_{0,2} $ & $ a50 $ \\ 
		& & $ (1- \gamma )^2\dfrac{1}{3} z_{0}z_{0,1} + \gamma (1- \gamma ) z_{0} \left(1-\dfrac{1}{3} z_{2,3}z_{1,3} \right) + \gamma ^2\dfrac{1}{3} z_{0}z_{0,2} $ & $ a51 $ \\ \hline
		$ (2,1,0,1) $ & $ 1 $ & $ (1- \gamma )^2\left(1-\dfrac{2}{3} z_{0}z_{0,1} \right) +2 \gamma (1- \gamma )\left(1- z_{0} +\dfrac{1}{3} z_{0}z_{1,3} \right) + \gamma ^2\left(1-\dfrac{2}{3} z_{0}z_{0,2}z_{2,3} \right) $ & $ a52 $ \\ 
		& & $ (1- \gamma )^2\dfrac{1}{3} z_{0}z_{0,1} + \gamma (1- \gamma ) z_{0} \left(1-\dfrac{1}{3} z_{1,3} \right) + \gamma ^2\dfrac{1}{3} z_{0}z_{0,2}z_{2,3} $ & $ a53 $ \\ 
		& & $ (1- \gamma )^2\dfrac{1}{3} z_{0}z_{0,1} + \gamma (1- \gamma ) z_{0} \left(1-\dfrac{1}{3} z_{1,3} \right) + \gamma ^2\dfrac{1}{3} z_{0}z_{0,2}z_{2,3} $ & $ a54 $ \\ \hline
		$ (2,2,0,0) $ & $ 4 $ & $ (1- \gamma )^2\left(1-\dfrac{2}{3} z_{1}z_{0}z_{0,1} \right) +2 \gamma (1- \gamma )\left(1-\dfrac{2}{3} z_{1}z_{0} \right) + \gamma ^2\left(1-\dfrac{2}{3} z_{1}z_{0}z_{0,2}z_{2,3}z_{1,3} \right) $ & $ a55 $ \\ 
		& & $ (1- \gamma )^2\dfrac{1}{3} z_{1}z_{0}z_{0,1} +2 \gamma (1- \gamma )\dfrac{1}{3} z_{1}z_{0} + \gamma ^2\dfrac{1}{3} z_{1}z_{0}z_{0,2}z_{2,3}z_{1,3} $ & $ a56 $ \\ 
		& & $ (1- \gamma )^2\dfrac{1}{3} z_{1}z_{0}z_{0,1} +2 \gamma (1- \gamma )\dfrac{1}{3} z_{1}z_{0} + \gamma ^2\dfrac{1}{3} z_{1}z_{0}z_{0,2}z_{2,3}z_{1,3} $ & $ a57 $ \\ \hline
  		\caption{Concordance factor equations for all 4-taxon subsets that involve the hybridization cycle.\label{table_cf}}
\end{longtable}

For the mapping of observed CFs to $a_i$ values, we need to know to which of the three splits each of the $a_i$'s corresponds to. So, next, we elaborate on which specific split each of the $a_i$ corresponds to:

\subsubsection*{$n=(0,0,2,2)$}

\begin{itemize}
    \item Individuals $k_1,k_2 \in n_2$ and $l_1,l_2 \in n_3$
    \item $P(k_1,k_2|l_1,l_2) = 1-\dfrac{2}{3} z_{2}z_{2,3}z_{3} = a_1$
    \item $P(k_1,l_1|k_2,l_2) = \dfrac{1}{3} z_{2}z_{2,3}z_{3} = a_2$
    \item $P(k_1,l_2|k_2,l_1) = \dfrac{1}{3} z_{2}z_{2,3}z_{3} = a_3$
\end{itemize}

\subsubsection*{$n=(0,1,2,1)$}

\begin{itemize}
    \item Individuals $j_1 \in n_1$; $k_1,k_2 \in n_2$ and $l_1 \in n_3$
    \item $P(k_1,k_2|j_1,l_1) = 1-\dfrac{2}{3} z_{2}z_{2,3} = a_4$
    \item $P(k_1,j_1|k_2,l_1) = \dfrac{1}{3} z_{2}z_{2,3} = a_5$
    \item $P(k_1,l_1|k_2,j_1) = \dfrac{1}{3} z_{2}z_{2,3} = a_6$
\end{itemize}

\subsubsection*{$n=(0,1,1,2)$}

\begin{itemize}
    \item Individuals $j_1 \in n_1$; $k_1 \in n_2$ and $l_1, l_2 \in n_3$
    \item $P(l_1,l_2|j_1,k_1) = 1-\dfrac{2}{3} z_{3} = a_7$
    \item $P(l_1,j_1|l_2,k_1) = \dfrac{1}{3} z_{3} = a_8$
    \item $P(l_1,k_1|l_2,j_1) = \dfrac{1}{3} z_{3} = a_9$
\end{itemize}

\subsubsection*{$n=(0,2,2,0)$}

\begin{itemize}
    \item Individuals $j_1, j_2 \in n_1$; $k_1, k_2 \in n_2$
    \item $P(j_1,j_2|k_1,k_2) = 1-\dfrac{2}{3} z_{2}z_{2,3}z_{1,3}z_1 = a_{10}$
    \item $P(j_1,k_1|j_2,k_2) = \dfrac{1}{3} z_{2}z_{2,3}z_{1,3}z_1 = a_{11}$
    \item $P(j_1,k_2|j_2,k_1) = \dfrac{1}{3} z_{2}z_{2,3}z_{1,3}z_1 = a_{12}$
\end{itemize}

\subsubsection*{$n=(0,2,1,1)$}

\begin{itemize}
    \item Individuals $j_1, j_2 \in n_1$; $k_1 \in n_2$; $l_1 \in n_3$
    \item $P(j_1,j_2|k_1,l_1) = 1-\dfrac{2}{3} z_{1,3}z_1 = a_{13}$
    \item $P(j_1,k_1|j_2,l_1) = \dfrac{1}{3} z_{1,3}z_1 = a_{14}$
    \item $P(j_1,l_1|j_2,k_1) = \dfrac{1}{3} z_{1,3}z_1 = a_{15}$
\end{itemize}

\subsubsection*{$n=(0,2,0,2)$}

\begin{itemize}
    \item Individuals $j_1, j_2 \in n_1$; $l_1, l_2 \in n_3$
    \item $P(j_1,j_2|l_1,l_2) = 1-\dfrac{2}{3} z_3z_{1,3}z_1 = a_{16}$
    \item $P(j_1,l_1|j_2,l_2) = \dfrac{1}{3} z_3z_{1,3}z_1 = a_{17}$
    \item $P(j_1,l_2|j_2,l_1) = \dfrac{1}{3} z_3z_{1,3}z_1 = a_{18}$
\end{itemize}

\subsubsection*{$n=(1,0,2,1)$}

\begin{itemize}
    \item Individuals $i_1 \in n_0$; $k_1, k_2 \in n_2$; $l_1 \in n_3$
    \item $P(k_1,k_2|i_1,l_1) = (1- \gamma )\left(1-\dfrac{2}{3} z_{2,3}z_{2} \right) + \gamma \left(1-\dfrac{2}{3} z_{2} \right) = a_{19}$
    \item $P(k_1,i_1|k_2,l_1) = (1- \gamma )\dfrac{1}{3} z_{2,3}z_{2} + \gamma \dfrac{1}{3} z_{2} = a_{20}$
    \item $P(k_1,l_1|k_2,i_1) = (1- \gamma )\dfrac{1}{3} z_{2,3}z_{2} + \gamma \dfrac{1}{3} z_{2} = a_{21}$
\end{itemize}

\subsubsection*{$n=(1,0,1,2)$}

\begin{itemize}
    \item Individuals $i_1 \in n_0$; $k_1 \in n_2$; $l_1,l_2 \in n_3$
    \item $P(l_1,l_2|i_1,k_1) = (1- \gamma )\left(1-\dfrac{2}{3} z_{3} \right) + \gamma \left(1-\dfrac{2}{3} z_{2,3}z_3 \right) = a_{22}$
    \item $P(l_1,i_1|l_2,k_1) = (1- \gamma )\dfrac{1}{3} z_{3} + \gamma \dfrac{1}{3} z_{2,3}z_3 = a_{23}$
    \item $P(l_1,k_1|l_2,i_1) = (1- \gamma )\dfrac{1}{3} z_{3} + \gamma \dfrac{1}{3} z_{2,3}z_3 = a_{24}$
\end{itemize}

\subsubsection*{$n=(1,1,2,0)$}

\begin{itemize}
    \item Individuals $i_1 \in n_0$; $j_1 \in n_1$; $k_1,k_2 \in n_2$
    \item $P(k_1,k_2|i_1,j_1) = (1- \gamma )\left(1-\dfrac{2}{3} z_{1,3}z_{2,3}z_{2} \right) + \gamma \left(1-\dfrac{2}{3} z_2 \right) = a_{25}$
    \item $P(k_1,i_1|k_2,j_1) = (1- \gamma )\dfrac{1}{3} z_{1,3}z_{2,3}z_{2} + \gamma \dfrac{1}{3} z_2 = a_{26}$
    \item $P(k_1,j_1|k_2,i_1) = (1- \gamma )\dfrac{1}{3} z_{1,3}z_{2,3}z_{2} + \gamma \dfrac{1}{3} z_2 = a_{27}$
\end{itemize}

\subsubsection*{$n=(1,1,1,1)$}

\begin{itemize}
    \item Individuals $i_1 \in n_0$; $j_1 \in n_1$; $k_1 \in n_2$; $l_1 \in n_3$
    \item $P(i_1,j_1|k_1,l_1) = (1- \gamma )\left(1-\dfrac{2}{3} z_{1,3} \right) + \gamma \dfrac{2}{3} z_{2,3} = a_{28}$
    \item $P(i_1,k_1|j_1,l_1) = (1- \gamma )\dfrac{1}{3} z_{1,3} + \gamma \left(1- \dfrac{2}{3} z_{2,3} \right) = a_{29}$
    \item $P(i_1,l_1|j_1,k_1) = (1- \gamma )\dfrac{1}{3} z_{1,3} + \gamma \dfrac{1}{3} z_{2,3} = a_{30}$
\end{itemize}

\subsubsection*{$n=(1,1,0,2)$}

\begin{itemize}
    \item Individuals $i_1 \in n_0$; $j_1 \in n_1$; $l_1,l_2 \in n_3$
    \item $P(l_1,l_2|i_1,j_1) = (1- \gamma )\left(1-\dfrac{2}{3} z_{1,3}z_{3} \right) + \gamma \left(1-\dfrac{2}{3} z_{3} \right) = a_{31}$
    \item $P(l_1,i_1|l_2,j_1) = (1- \gamma )\dfrac{1}{3} z_{1,3}z_{3} + \gamma \dfrac{1}{3} z_{3} = a_{32}$
    \item $P(l_1,j_1|l_2,i_1) = (1- \gamma )\dfrac{1}{3} z_{1,3}z_{3} + \gamma \dfrac{1}{3} z_{3} = a_{33}$
\end{itemize}


\subsubsection*{$n=(1,2,1,0)$}
\begin{itemize}
    \item Individuals $i_1 \in n_0$; $j_1, j_2 \in n_1$; $k_1 \in n_2$
    \item $P(j_1,j_2|i_1,k_1) = (1- \gamma )\left(1-\dfrac{2}{3} z_{1} \right) + \gamma \left(1-\dfrac{2}{3} z_{2,3}z_{1,3}z_{1} \right) = a_{34}$
    \item $P(j_1,i_1|j_2,k_1) = (1- \gamma )\dfrac{1}{3} z_{1} + \gamma \dfrac{1}{3} z_{2,3}z_{1,3}z_{1} = a_{35}$
    \item $P(j_1,k_1|j_2,i_1) = (1- \gamma )\dfrac{1}{3} z_{1} + \gamma \dfrac{1}{3} z_{2,3}z_{1,3}z_{1} = a_{36}$
\end{itemize}

\subsubsection*{$n=(1,2,0,1)$}
\begin{itemize}
    \item Individuals $i_1 \in n_0$; $j_1, j_2 \in n_1$; $l_1 \in n_3$
    \item $P(j_1,j_2|i_1,l_1) = (1- \gamma )\left(1-\dfrac{2}{3} z_{1} \right) + \gamma \left(1-\dfrac{2}{3} z_{1,3}z_{1} \right) = a_{37}$
    \item $P(j_1,l_1|j_2,i_1) = (1- \gamma )\dfrac{1}{3} z_{1} + \gamma \dfrac{1}{3} z_{1,3}z_{1} = a_{38}$
    \item $P(j_1,i_1|j_2,l_1) = (1- \gamma )\dfrac{1}{3} z_{1} + \gamma \dfrac{1}{3} z_{1,3}z_{1} = a_{39}$
\end{itemize}

\subsubsection*{$n=(2,0,2,0)$}
\begin{itemize}
    \item Individuals $i_1, i_2 \in n_0$; $k_1, k_2 \in n_2$
    \item $P(i_1,i_2|k_1,k_2) = (1- \gamma )^2\left(1-\dfrac{2}{3} z_{2}z_{0}z_{0,1}z_{1,3}z_{2,3} \right) +2 \gamma (1- \gamma )\left(1-\dfrac{2}{3} z_{2}z_{0} \right) + \gamma ^2\left(1-\dfrac{2}{3} z_{2}z_{0}z_{0,2} \right) = a_{40}$
    \item $P(i_1,k_1|i_2,k_2) = (1- \gamma )^2\dfrac{1}{3} z_{2}z_{0}z_{0,1}z_{1,3}z_{2,3} +2 \gamma (1- \gamma )\dfrac{1}{3} z_{2}z_{0} + \gamma ^2\dfrac{1}{3} z_{2}z_{0}z_{0,2} = a_{41}$
    \item $P(i_1,k_2|k_1,i_2) = (1- \gamma )^2\dfrac{1}{3} z_{2}z_{0}z_{0,1}z_{1,3}z_{2,3} +2 \gamma (1- \gamma )\dfrac{1}{3} z_{2}z_{0} + \gamma ^2\dfrac{1}{3} z_{2}z_{0}z_{0,2} = a_{42}$
\end{itemize}

\subsubsection*{$n=(2,0,1,1)$}
\begin{itemize}
    \item Individuals $i_1, i_2 \in n_0$; $k_1 \in n_2$; $l_1 \in n_3$
    \item $P(i_1,i_2|k_1,l_1) = (1- \gamma )^2\left(1-\dfrac{2}{3} z_{0}z_{1,3}z_{0,1} \right) +2 \gamma (1- \gamma )\left(1- z_{0} +\dfrac{1}{3} z_{0}z_{2,3} \right) + \gamma ^2\left(1-\dfrac{2}{3} z_{0}z_{0,2} \right) = a_{43}$
    \item $P(i_1,k_1|l_1,i_2) = (1- \gamma )^2\dfrac{1}{3} z_{0}z_{1,3}z_{0,1} + \gamma (1- \gamma ) z_{0} \left(1-\dfrac{1}{3} z_{2,3} \right) + \gamma ^2\dfrac{1}{3} z_{0}z_{0,2} = a_{44}$
    \item $P(i_1,l_1|i_2,k_1) = (1- \gamma )^2\dfrac{1}{3} z_{0}z_{1,3}z_{0,1} + \gamma (1- \gamma ) z_{0} \left(1-\dfrac{1}{3} z_{2,3} \right) + \gamma ^2\dfrac{1}{3} z_{0}z_{0,2} = a_{45}$
\end{itemize}

\subsubsection*{$n=(2,0,0,2)$}
\begin{itemize}
    \item Individuals $i_1, i_2 \in n_0$; $l_1, l_2 \in n_3$
    \item $P(i_1,i_2|l_1,l_2) = (1- \gamma )^2\left(1-\dfrac{2}{3} z_{3}z_{0}z_{1,3}z_{0,1} \right) +2 \gamma (1- \gamma )\left(1-\dfrac{2}{3} z_{3}z_{0} \right) + \gamma ^2\left(1-\dfrac{2}{3} z_{3}z_{0}z_{2,3}z_{0,2} \right) = a_{46}$
    \item $P(i_1,l_1|i_2,l_2) = (1- \gamma )^2\dfrac{1}{3} z_{3}z_{0}z_{1,3}z_{0,1} +2 \gamma (1- \gamma )\dfrac{1}{3} z_{3}z_{0} + \gamma ^2\dfrac{1}{3} z_{3}z_{0}z_{2,3}z_{0,2} = a_{47}$
    \item $P(i_1,l_2|i_2,l_1) = (1- \gamma )^2\dfrac{1}{3} z_{3}z_{0}z_{1,3}z_{0,1} +2 \gamma (1- \gamma )\dfrac{1}{3} z_{3}z_{0} + \gamma ^2\dfrac{1}{3} z_{3}z_{0}z_{2,3}z_{0,2} = a_{48}$
\end{itemize}

\subsubsection*{$n=(2,1,1,0)$}
\begin{itemize}
    \item Individuals $i_1, i_2 \in n_0$; $j_1 \in n_1$; $k_1 \in n_2$
    \item $P(i_1,i_2|j_1,k_1) = (1- \gamma )^2\left(1-\dfrac{2}{3} z_{0}z_{0,1} \right) +2 \gamma (1- \gamma )\left(1- z_{0} +\dfrac{1}{3} z_{0}z_{2,3}z_{1,3} \right) + \gamma ^2\left(1-\dfrac{2}{3} z_{0}z_{0,2} \right) = a_{49}$
    \item $P(i_1,j_1|i_2,k_1) = 1- \gamma )^2\dfrac{1}{3} z_{0}z_{0,1} + \gamma (1- \gamma ) z_{0} \left(1-\dfrac{1}{3} z_{2,3}z_{1,3} \right) + \gamma ^2\dfrac{1}{3} z_{0}z_{0,2} = a_{50}$
    \item $P(i_1,k_1|i_2,j_1) = 1- \gamma )^2\dfrac{1}{3} z_{0}z_{0,1} + \gamma (1- \gamma ) z_{0} \left(1-\dfrac{1}{3} z_{2,3}z_{1,3} \right) + \gamma ^2\dfrac{1}{3} z_{0}z_{0,2} = a_{51}$
\end{itemize}

\subsubsection*{$n=(2,1,0,1)$}
\begin{itemize}
    \item Individuals $i_1, i_2 \in n_0$; $j_1 \in n_1$; $l_1 \in n_3$
    \item $P(i_1,i_2|j_1,l_1) = (1- \gamma )^2\left(1-\dfrac{2}{3} z_{0}z_{0,1} \right) +2 \gamma (1- \gamma )\left(1- z_{0} +\dfrac{1}{3} z_{0}z_{1,3} \right) + \gamma ^2\left(1-\dfrac{2}{3} z_{0}z_{0,2}z_{2,3} \right) = a_{52}$
    \item $P(i_1,j_1|i_2,l_1) = (1- \gamma )^2\dfrac{1}{3} z_{0}z_{0,1} + \gamma (1- \gamma ) z_{0} \left(1-\dfrac{1}{3} z_{1,3} \right) + \gamma ^2\dfrac{1}{3} z_{0}z_{0,2}z_{2,3} = a_{53}$
    \item $P(i_1,l_1|i_2,j_1) = (1- \gamma )^2\dfrac{1}{3} z_{0}z_{0,1} + \gamma (1- \gamma ) z_{0} \left(1-\dfrac{1}{3} z_{1,3} \right) + \gamma ^2\dfrac{1}{3} z_{0}z_{0,2}z_{2,3} = a_{54}$
\end{itemize}

\subsubsection*{$n=(2,2,0,0)$}
\begin{itemize}
    \item Individuals $i_1, i_2 \in n_0$; $j_1, j_2 \in n_1$; 
    \item $P(i_1,i_2|j_1,j_2) = (1- \gamma )^2\left(1-\dfrac{2}{3} z_{1}z_{0}z_{0,1} \right) +2 \gamma (1- \gamma )\left(1-\dfrac{2}{3} z_{1}z_{0} \right) + \gamma ^2\left(1-\dfrac{2}{3} z_{1}z_{0}z_{0,2}z_{2,3}z_{1,3} \right) = a_{55}$
    \item $P(i_1,j_1|i_2,j_2) = (1- \gamma )^2\dfrac{1}{3} z_{1}z_{0}z_{0,1} +2 \gamma (1- \gamma )\dfrac{1}{3} z_{1}z_{0} + \gamma ^2\dfrac{1}{3} z_{1}z_{0}z_{0,2}z_{2,3}z_{1,3} = a_{56}$
    \item $P(i_1,j_2|i_2,j_1) = (1- \gamma )^2\dfrac{1}{3} z_{1}z_{0}z_{0,1} +2 \gamma (1- \gamma )\dfrac{1}{3} z_{1}z_{0} + \gamma ^2\dfrac{1}{3} z_{1}z_{0}z_{0,2}z_{2,3}z_{1,3} = a_{57}$
\end{itemize}

\section{Phylogenetic invariants for $n$-taxon phylogenetic networks with one 4-node hybridization cycle}
\label{invs}

Below, we present the invariants for different networks $N$ all with one 4-cycle, but with different number of species on the clades $n_0,n_1,n_2,n_3$. For example, the network $N=1112$ corresponds to a network with 5 species: one in $n_0$, one in $n_1$, one in $n_2$ and two in $n_3$. The number of species defines the number of CF formulas. For example, for 6 species, there are ${6 \choose 4} = 15$ 4-taxon subsets, each with 3 CF formulas. So, for 6 species, we have 45 CF formulas and thus, 45 CF values. However, we only want to focus on the 4-taxon subsets that involve the hybridization cycle. For the case of $N=1112$, they are only 4: $(0,1,1,2), (1,0,1,2), (1,1,1,1), (1,1,0,2)$.
Note that our table of CF formulas has 57 different CF formulas (and therefore, values). This discrepancy is due to the fact that the table is listing all possible CF formulas and we will have fewer formulas if we have less than 8 species.

For some examples, for computational restrictions, we had to include just a subset of the original CF equations to obtain the Gr\"{o}bner basis in $a_i$. We denote these cases with "subset". All Macaulay2 scripts (and output) can be found in the GitHub repository: \url{https://github.com/solislemuslab/phylo-diamond.jl}.

\begin{itemize}
		\item $ N =1112$      
		\begin{enumerate}
			\item $
a_{32} - a_{33}$
\item $
 a_{31} + 2a_{33} - 1$
\item $
 a_{28} + a_{29} + a_{30} - 1$
\item $
 a_{23} - a_{24}$
\item $
 a_{22} + 2a_{24} - 1$
\item $
 a_{8} - a_{9}$
\item $
 a_{7} + 2a_{9} - 1$
\item $
 3a_{9}*a_{30} + a_{9} - a_{24} - a_{33}$
\item $
 a_{24}*a_{29} + 2a_{24}*a_{30} + a_{29}*a_{33} - a_{30}*a_{33} - a_{33}$
\item $
 3a_{9}*a_{29} - 2a_{9} + 2a_{24} - a_{33}$
		\end{enumerate}
		
		\item $ N =1121$      
		\begin{enumerate}
			\item $
a_{28} + a_{29} + a_{30} - 1$
\item $
 a_{26} - a_{27}$
\item $
 a_{25} + 2a_{27} - 1$
\item $
 a_{20} - a_{21}$
\item $
 a_{19} + 2a_{21} - 1$
\item $
 a_{5} - a_{6}$
\item $
 a_{4} + 2a_{6} - 1$
\item $
 a_{6}*a_{29} + 2a_{6}*a_{30} - a_{6} + a_{21} - a_{27}$
		\end{enumerate}
		
		\item $ N =1122$      
		\begin{enumerate}
			\item $
a_{32} - a_{33}$
\item $
 a_{31} + 2a_{33} - 1$
\item $
 a_{28} + a_{29} + a_{30} - 1$
\item $
 a_{26} - a_{27}$
\item $
 a_{25} + 2a_{27} - 1$
\item $
 a_{23} - a_{24}$
\item $
 a_{22} + 2a_{24} - 1$
\item $
 a_{20} - a_{21}$
\item $
 a_{19} + 2a_{21} - 1$
\item $
 a_{8} - a_{9}$
\item $
 a_{7} + 2a_{9} - 1$
\item $
 a_{5} - a_{6}$
\item $
 a_{4} + 2a_{6} - 1$
\item $
 a_{2} - a_{3}$
\item $
 a_{1} + 2a_{3} - 1$
\item $
 3a_{9}*a_{30} + a_{9} - a_{24} - a_{33}$
\item $
 a_{24}*a_{29} + 2a_{24}*a_{30} + a_{29}*a_{33} - a_{30}*a_{33} - a_{33}$
\item $
 3a_{9}*a_{29} - 2a_{9} + 2a_{24} - a_{33}$
\item $
 a_{6}*a_{29} + 2a_{6}*a_{30} - a_{6} + a_{21} - a_{27}$
\item $
 a_{3}*a_{29} + 2a_{3}*a_{30} - 3a_{6}*a_{33}$
\item $
 3a_{6}*a_{24} - 3a_{3}*a_{30} + 3a_{6}*a_{33} - a_{3}$
\item $
 3a_{9}*a_{21} - 3a_{9}*a_{27} + 3a_{6}*a_{33} - a_{3}$
\item $
 3a_{6}*a_{9} - a_{3}$
		\end{enumerate}
		
		\item $ N =1211$      
		\begin{enumerate}
			\item $
a_{38} - a_{39}$
\item $
 a_{37} + 2a_{39} - 1$
\item $
 a_{35} - a_{36}$
\item $
 a_{34} + 2a_{36} - 1$
\item $
 a_{28} + a_{29} + a_{30} - 1$
\item $
 a_{14} - a_{15}$
\item $
 a_{13} + 2a_{15} - 1$
\item $
 a_{15}*a_{29} - a_{15}*a_{30} + a_{36} - a_{39}$
		\end{enumerate}
		
		\item $ N =1212$      
		\begin{enumerate}
			\item $
a_{38} - a_{39}$
\item $
 a_{37} + 2a_{39} - 1$
\item $
 a_{35} - a_{36}$
\item $
 a_{34} + 2a_{36} - 1$
\item $
 a_{32} - a_{33}$
\item $
 a_{31} + 2a_{33} - 1$
\item $
 a_{28} + a_{29} + a_{30} - 1$
\item $
 a_{23} - a_{24}$
\item $
 a_{22} + 2a_{24} - 1$
\item $
 a_{17} - a_{18}$
\item $
 a_{16} + 2a_{18} - 1$
\item $
 a_{14} - a_{15}$
\item $
 a_{13} + 2a_{15} - 1$
\item $
 a_{8} - a_{9}$
\item $
 a_{7} + 2a_{9} - 1$
\item $
 a_{18}*a_{30} - a_{15}*a_{33} - a_{9}*a_{36} + a_{9}*a_{39}$
\item $
 3a_{9}*a_{30} + a_{9} - a_{24} - a_{33}$
\item $
 a_{24}*a_{29} + 2a_{24}*a_{30} + a_{29}*a_{33} - a_{30}*a_{33} - a_{33}$
\item $
 a_{18}*a_{29} - a_{15}*a_{33} + 2a_{9}*a_{36} - 2a_{9}*a_{39}$
\item $
 a_{15}*a_{29} - a_{15}*a_{30} + a_{36} - a_{39}$
\item $
 3a_{9}*a_{29} - 2a_{9} + 2a_{24} - a_{33}$
\item $
 3a_{15}*a_{24} - 3a_{9}*a_{36} + 3a_{9}*a_{39} - a_{18}$
\item $
 3a_{9}*a_{15} - a_{18}$ 
		\end{enumerate}
		
		\item $ N =1221$      
		\begin{enumerate}
			\item $
a_{38} - a_{39}$
\item $
 a_{37} + 2a_{39} - 1$
\item $
 a_{35} - a_{36}$
\item $
 a_{34} + 2a_{36} - 1$
\item $
 a_{28} + a_{29} + a_{30} - 1$
\item $
 a_{26} - a_{27}$
\item $
 a_{25} + 2a_{27} - 1$
\item $
 a_{20} - a_{21}$
\item $
 a_{19} + 2a_{21} - 1$
\item $
 a_{14} - a_{15}$
\item $
 a_{13} + 2a_{15} - 1$
\item $
 a_{11} - a_{12}$
\item $
 a_{10} + 2a_{12} - 1$
\item $
 a_{5} - a_{6}$
\item $
 a_{4} + 2a_{6} - 1$
\item $
 a_{15}*a_{29} - a_{15}*a_{30} + a_{36} - a_{39}$
\item $
 a_{12}*a_{29} - a_{12}*a_{30} + 3a_{6}*a_{36} - 3a_{6}*a_{39}$
\item $
 a_{6}*a_{29} + 2a_{6}*a_{30} - a_{6} + a_{21} - a_{27}$
\item $
 3a_{15}*a_{21} - 3a_{15}*a_{27} + 3a_{12}*a_{30} - 3a_{6}*a_{36} + 3a_{6}*a_{39} - a_{12}$
\item $
 3a_{6}*a_{15} - a_{12}$
\item $
 a_{21}*a_{29}*a_{36} + 2a_{21}*a_{30}*a_{36} - a_{27}*a_{29}*a_{39} + a_{27}*a_{30}*a_{39} - a_{15}*a_{27} + a_{12}*a_{30} - a_{6}*a_{36} - a_{21}*a_{36} - a_{27}*a_{36} + 2a_{21}*a_{39}$
		\end{enumerate}
		
		\item $ N =1222$      
		\begin{enumerate}
			\item $
a_{38} - a_{39}$
\item $
 a_{37} + 2a_{39} - 1$
\item $
 a_{35} - a_{36}$
\item $
 a_{34} + 2a_{36} - 1$
\item $
 a_{32} - a_{33}$
\item $
 a_{31} + 2a_{33} - 1$
\item $
 a_{28} + a_{29} + a_{30} - 1$
\item $
 a_{26} - a_{27}$
\item $
 a_{25} + 2a_{27} - 1$
\item $
 a_{23} - a_{24}$
\item $
 a_{22} + 2a_{24} - 1$
\item $
 a_{20} - a_{21}$
\item $
 a_{19} + 2a_{21} - 1$
\item $
 a_{17} - a_{18}$
\item $
 a_{16} + 2a_{18} - 1$
\item $
 a_{14} - a_{15}$
\item $
 a_{13} + 2a_{15} - 1$
\item $
 a_{11} - a_{12}$
\item $
 a_{10} + 2a_{12} - 1$
\item $
 a_{8} - a_{9}$
\item $
 a_{7} + 2a_{9} - 1$
\item $
 a_{5} - a_{6}$
\item $
 a_{4} + 2a_{6} - 1$
\item $
 a_{2} - a_{3}$
\item $
 a_{1} + 2a_{3} - 1$
\item $
 a_{18}*a_{30} - a_{15}*a_{33} - a_{9}*a_{36} + a_{9}*a_{39}$
\item $
 3a_{9}*a_{30} + a_{9} - a_{24} - a_{33}$
\item $
 a_{24}*a_{29} + 2a_{24}*a_{30} + a_{29}*a_{33} - a_{30}*a_{33} - a_{33}$
\item $
 a_{18}*a_{29} - a_{15}*a_{33} + 2a_{9}*a_{36} - 2a_{9}*a_{39}$
\item $
 a_{15}*a_{29} - a_{15}*a_{30} + a_{36} - a_{39}$
\item $
 a_{12}*a_{29} - a_{12}*a_{30} + 3a_{6}*a_{36} - 3a_{6}*a_{39}$
\item $
 3a_{9}*a_{29} - 2a_{9} + 2a_{24} - a_{33}$
\item $
 a_{6}*a_{29} + 2a_{6}*a_{30} - a_{6} + a_{21} - a_{27}$
\item $
 a_{3}*a_{29} + 2a_{3}*a_{30} - 3a_{6}*a_{33}$
\item $
 3a_{15}*a_{24} - 3a_{9}*a_{36} + 3a_{9}*a_{39} - a_{18}$
\item $
 3a_{6}*a_{24} - 3a_{3}*a_{30} + 3a_{6}*a_{33} - a_{3}$
\item $
 a_{18}*a_{21} - a_{12}*a_{24} - a_{18}*a_{27} + a_{12}*a_{33} + a_{3}*a_{36} - a_{3}*a_{39}$
\item $
 3a_{15}*a_{21} - 3a_{15}*a_{27} + 3a_{12}*a_{30} - 3a_{6}*a_{36} + 3a_{6}*a_{39} - a_{12}$
\item $
 3a_{9}*a_{21} - 3a_{9}*a_{27} + 3a_{6}*a_{33} - a_{3}$
\item $
 a_{6}*a_{18} - a_{12}*a_{24} + a_{3}*a_{36} - a_{3}*a_{39}$
\item $
 3a_{9}*a_{15} - a_{18}$
\item $
 3a_{6}*a_{15} - a_{12}$
\item $
 a_{3}*a_{15} - a_{12}*a_{24} + a_{3}*a_{36} - a_{3}*a_{39}$
\item $
 a_{9}*a_{12} - a_{12}*a_{24} + a_{3}*a_{36} - a_{3}*a_{39}$
\item $
 3a_{6}*a_{9} - a_{3}$
\item $
 a_{21}*a_{29}*a_{36} + 2a_{21}*a_{30}*a_{36} - a_{27}*a_{29}*a_{39} + a_{27}*a_{30}*a_{39} - a_{15}*a_{27} + a_{12}*a_{30} - a_{6}*a_{36} - a_{21}*a_{36} - a_{27}*a_{36} + 2a_{21}*a_{39}$
\item $
 6a_{9}*a_{27}*a_{36} - 3a_{6}*a_{33}*a_{36} - 3a_{21}*a_{33}*a_{36} - 3a_{9}*a_{27}*a_{39} - 3a_{24}*a_{27}*a_{39} + 6a_{6}*a_{33}*a_{39} + a_{18}*a_{27} - a_{12}*a_{33} + a_{3}*a_{36} - a_{3}*a_{39}$ 
		\end{enumerate}
		
		\item $ N =2111$      
		\begin{enumerate}
			\item $
a_{53} - a_{54}$
\item $
 a_{52} + 2a_{54} - 1$
\item $
 a_{50} - a_{51}$
\item $
 a_{49} + 2a_{51} - 1$
\item $
 a_{44} - a_{45}$
\item $
 a_{43} + 2a_{45} - 1$
\item $
 a_{28} + a_{29} + a_{30} - 1$
		\end{enumerate}
		
		\item $ N =2112$      
		\begin{enumerate}
			\item $
a_{53} - a_{54}$
\item $
 a_{52} + 2a_{54} - 1$
\item $
 a_{50} - a_{51}$
\item $
 a_{49} + 2a_{51} - 1$
\item $
 a_{47} - a_{48}$
\item $
 a_{46} + 2a_{48} - 1$
\item $
 a_{44} - a_{45}$
\item $
 a_{43} + 2a_{45} - 1$
\item $
 a_{32} - a_{33}$
\item $
 a_{31} + 2a_{33} - 1$
\item $
 a_{28} + a_{29} + a_{30} - 1$
\item $
 a_{23} - a_{24}$
\item $
 a_{22} + 2a_{24} - 1$
\item $
 a_{8} - a_{9}$
\item $
 a_{7} + 2a_{9} - 1$
\item $
 3a_{9}*a_{30} + a_{9} - a_{24} - a_{33}$
\item $
 a_{24}*a_{29} + 2a_{24}*a_{30} + a_{29}*a_{33} - a_{30}*a_{33} - a_{33}$
\item $
 3a_{9}*a_{29} - 2a_{9} + 2a_{24} - a_{33}$
		\end{enumerate}

		\item $ N =2121$
		\begin{enumerate}
		\item $
a_{4} + 2a_{6} - 1$
\item $
 a_{5} - a_{6}$
\item $
 a_{28} + a_{29} + a_{30} - 1$
\item $
 a_{19} + 2a_{21} - 1$
\item $
 a_{20} - a_{21}$
\item $
 a_{25} + 2a_{27} - 1$
\item $
 a_{26} - a_{27}$
\item $
 a_{40} + 2a_{42} - 1$
\item $
 a_{41} - a_{42}$
\item $
 a_{43} + 2a_{45} - 1$
\item $
 a_{44} - a_{45}$
\item $
 a_{49} + 2a_{51} - 1$
\item $
 a_{50} - a_{51}$
\item $
 a_{52} + 2a_{54} - 1$
\item $
 a_{53} - a_{54}$
			\item $ a_{5}*a_{29}+2a_{5}*a_{30}-a_{5}+a_{20}-a_{26} $ 
			\item $ 2a_{20}*a_{29}^{3}a_{41}+a_{26}*a_{29}^{3}a_{41}+3a_{20}*a_{29}^{2}a_{30}*a_{41}-3a_{20}*a_{29}*a_{30}^{2}a_{41}-3a_{26}*a_{29}*a_{30}^{2}a_{41}-2a_{20}*a_{30}^{3}a_{41}+2a_{26}*a_{30}^{3}a_{41}-3a_{20}*a_{26}*a_{29}^{2}a_{44}-3a_{26}^{2}a_{29}^{2}a_{44}-3a_{20}*a_{26}*a_{29}*a_{30}*a_{44}+6a_{26}^{2}a_{29}*a_{30}*a_{44}+6a_{20}*a_{26}*a_{30}^{2}a_{44}-3a_{26}^{2}a_{30}^{2}a_{44}-6a_{20}^{2}a_{29}^{2}a_{50}+3a_{20}*a_{26}*a_{29}^{2}a_{50}-15a_{20}^{2}a_{29}*a_{30}*a_{50}-6a_{20}*a_{26}*a_{29}*a_{30}*a_{50}-6a_{20}^{2}a_{30}^{2}a_{50}+3a_{20}*a_{26}*a_{30}^{2}a_{50}+3a_{20}*a_{26}*a_{29}^{2}a_{53}+3a_{20}*a_{26}*a_{29}*a_{30}*a_{53}-6a_{20}*a_{26}*a_{30}^{2}a_{53}-4a_{20}*a_{29}^{2}a_{41}-2a_{26}*a_{29}^{2}a_{41}-a_{20}*a_{29}*a_{30}*a_{41}+a_{26}*a_{29}*a_{30}*a_{41}+5a_{20}*a_{30}^{2}a_{41}+a_{26}*a_{30}^{2}a_{41}+3a_{20}^{2}a_{29}*a_{44}+6a_{20}*a_{26}*a_{29}*a_{44}+3a_{26}^{2}a_{29}*a_{44}+6a_{20}^{2}a_{30}*a_{44}+18a_{5}*a_{26}*a_{30}*a_{44}-6a_{20}*a_{26}*a_{30}*a_{44}-3a_{26}^{2}a_{30}*a_{44}+6a_{20}*a_{26}*a_{29}*a_{50}-9a_{5}*a_{20}*a_{30}*a_{50}+9a_{20}^{2}a_{30}*a_{50}-9a_{5}*a_{26}*a_{30}*a_{50}+12a_{20}*a_{26}*a_{30}*a_{50}-3a_{20}^{2}a_{29}*a_{53}-3a_{26}^{2}a_{29}*a_{53}-6a_{20}^{2}a_{30}*a_{53}+3a_{26}^{2}a_{30}*a_{53}+3a_{26}*a_{29}*a_{41}-3a_{26}*a_{30}*a_{41}-3a_{5}^{2}a_{44}+3a_{5}*a_{20}*a_{44}-3a_{5}*a_{26}*a_{44}-6a_{26}^{2}a_{44}+3a_{5}^{2}a_{50}-6a_{5}*a_{20}*a_{50}+3a_{20}^{2}a_{50}+6a_{5}*a_{26}*a_{50}-6a_{20}*a_{26}*a_{50}+3a_{5}*a_{20}*a_{53}-3a_{20}^{2}a_{53}-3a_{5}*a_{26}*a_{53}+6a_{20}*a_{26}*a_{53} $ 
		\end{enumerate}

		\item $ N =2211$ 
		\begin{enumerate}
		\item $
a_{13} + 2a_{15} - 1$
\item $
 a_{14} - a_{15}$
\item $
 a_{28} + a_{29} + a_{30} - 1$
\item $
 a_{34} + 2a_{36} - 1$
\item $
 a_{35} - a_{36}$
\item $
 a_{37} + 2a_{39} - 1$
\item $
 a_{38} - a_{39}$
\item $
 a_{43} + 2a_{45} - 1$
\item $
 a_{44} - a_{45}$
\item $
 a_{49} + 2a_{51} - 1$
\item $
 a_{50} - a_{51}$
\item $
 a_{52} + 2a_{54} - 1$
\item $
 a_{53} - a_{54}$
\item $
 a_{55} + 2a_{57} - 1$
\item $
 a_{56} - a_{57}$
			\item $ a_{14}*a_{29}-a_{14}*a_{30}+a_{35}-a_{38} $ 
			\item $ 3a_{29}^{2}a_{35}*a_{38}*a_{44}+3a_{29}*a_{30}*a_{35}*a_{38}*a_{44}-6a_{30}^{2}a_{35}*a_{38}*a_{44}+3a_{29}^{2}a_{35}*a_{38}*a_{50}+12a_{29}*a_{30}*a_{35}*a_{38}*a_{50}+12a_{30}^{2}a_{35}*a_{38}*a_{50}-6a_{29}^{2}a_{38}^{2}a_{50}+3a_{29}*a_{30}*a_{38}^{2}a_{50}+3a_{30}^{2}a_{38}^{2}a_{50}-3a_{29}^{2}a_{35}^{2}a_{53}-12a_{29}*a_{30}*a_{35}^{2}a_{53}-12a_{30}^{2}a_{35}^{2}a_{53}-3a_{29}^{2}a_{35}*a_{38}*a_{53}-3a_{29}*a_{30}*a_{35}*a_{38}*a_{53}+6a_{30}^{2}a_{35}*a_{38}*a_{53}-a_{29}^{3}a_{35}*a_{56}-3a_{29}^{2}a_{30}*a_{35}*a_{56}+4a_{30}^{3}a_{35}*a_{56}-2a_{29}^{3}a_{38}*a_{56}-3a_{29}^{2}a_{30}*a_{38}*a_{56}+3a_{29}*a_{30}^{2}a_{38}*a_{56}+2a_{30}^{3}a_{38}*a_{56}+3a_{29}*a_{35}^{2}a_{44}+6a_{30}*a_{35}^{2}a_{44}-6a_{29}*a_{35}*a_{38}*a_{44}-3a_{30}*a_{35}*a_{38}*a_{44}+3a_{29}*a_{38}^{2}a_{44}-3a_{30}*a_{38}^{2}a_{44}-9a_{14}*a_{30}*a_{35}*a_{50}-9a_{14}*a_{30}*a_{38}*a_{50}-12a_{29}*a_{35}*a_{38}*a_{50}-6a_{30}*a_{35}*a_{38}*a_{50}+12a_{29}*a_{38}^{2}a_{50}+6a_{30}*a_{38}^{2}a_{50}+18a_{14}*a_{30}*a_{35}*a_{53}+3a_{29}*a_{35}^{2}a_{53}+6a_{30}*a_{35}^{2}a_{53}-9a_{30}*a_{35}*a_{38}*a_{53}-3a_{29}*a_{38}^{2}a_{53}+3a_{30}*a_{38}^{2}a_{53}+a_{29}^{2}a_{35}*a_{56}+a_{29}*a_{30}*a_{35}*a_{56}-2a_{30}^{2}a_{35}*a_{56}+2a_{29}^{2}a_{38}*a_{56}-a_{29}*a_{30}*a_{38}*a_{56}-a_{30}^{2}a_{38}*a_{56}-3a_{14}*a_{35}*a_{44}-3a_{35}^{2}a_{44}+3a_{14}*a_{38}*a_{44}+9a_{35}*a_{38}*a_{44}-6a_{38}^{2}a_{44}+3a_{14}^{2}a_{50}+6a_{14}*a_{35}*a_{50}-6a_{14}*a_{38}*a_{50}+3a_{35}*a_{38}*a_{50}-3a_{38}^{2}a_{50}-3a_{14}^{2}a_{53}-3a_{14}*a_{35}*a_{53}-6a_{35}^{2}a_{53}+3a_{14}*a_{38}*a_{53}+3a_{35}*a_{38}*a_{53}+3a_{38}^{2}a_{53}-2a_{29}*a_{35}*a_{56}-4a_{30}*a_{35}*a_{56}+2a_{29}*a_{38}*a_{56}+4a_{30}*a_{38}*a_{56}+2a_{35}*a_{56}-2a_{38}*a_{56} $ 
		\end{enumerate}

		\item $ N =2212$ 
\begin{enumerate}
\item $
a_{16} + 2a_{18} - 1$
\item $
 a_{17} - a_{18}$
\item $
 a_{28} + a_{29} + a_{30} - 1$
\item $
 a_{13} + 2a_{15} - 1$
\item $
 a_{14} - a_{15}$
\item $
 a_{31} + 2a_{33} - 1$
\item $
 a_{32} - a_{33}$
\item $
 a_{7} + 2a_{9} - 1$
\item $
 a_{8} - a_{9}$
\item $
 a_{34} + 2a_{36} - 1$
\item $
 a_{35} - a_{36}$
\item $
 a_{37} + 2a_{39} - 1$
\item $
 a_{38} - a_{39}$
\item $
 a_{22} + 2a_{24} - 1$
\item $
 a_{23} - a_{24}$
\item $
 a_{43} + 2a_{45} - 1$
\item $
 a_{44} - a_{45}$
\item $
 a_{46} + 2a_{48} - 1$
\item $
 a_{47} - a_{48}$
\item $
 a_{49} + 2a_{51} - 1$
\item $
 a_{50} - a_{51}$
\item $
 a_{52} + 2a_{54} - 1$
\item $
 a_{53} - a_{54}$
\item $
 a_{55} + 2a_{57} - 1$
\item $
 a_{56} - a_{57}$
\item $
a_{17}*a_{30} - a_{14}*a_{32} - a_{8}*a_{35} + a_{8}*a_{38}$
\item $
 3a_{8}*a_{30} + a_{8} - a_{23} - a_{32}$
\item $
 a_{23}*a_{29} + 2a_{23}*a_{30} + a_{29}*a_{32} - a_{30}*a_{32} - a_{32}$
\item $
 a_{17}*a_{29} - a_{14}*a_{32} + 2a_{8}*a_{35} - 2a_{8}*a_{38}$
\item $
 a_{14}*a_{29} - a_{14}*a_{30} + a_{35} - a_{38}$
\item $
 3a_{8}*a_{29} - 2a_{8} + 2a_{23} - a_{32}$
\item $
 3a_{14}*a_{23} - 3a_{8}*a_{35} + 3a_{8}*a_{38} - a_{17}$
\item $
 3a_{8}*a_{14} - a_{17}$
\item $
 3a_{32}*a_{38}*a_{44} - 2a_{29}*a_{38}*a_{47} - a_{30}*a_{38}*a_{47} - 3a_{32}*a_{38}*a_{50} + 3a_{32}*a_{35}*a_{53} + 3a_{8}*a_{38}*a_{53} - 3a_{23}*a_{38}*a_{53} + a_{29}*a_{32}*a_{56} - a_{30}*a_{32}*a_{56} - a_{17}*a_{44} + a_{14}*a_{47} - a_{35}*a_{47} + a_{38}*a_{47} + a_{17}*a_{50} - a_{17}*a_{53} - a_{8}*a_{56} + a_{23}*a_{56}$
\item $
 3a_{8}*a_{35}*a_{44} - 3a_{32}*a_{35}*a_{44} - 3a_{8}*a_{38}*a_{44} + a_{29}*a_{35}*a_{47} + 2a_{30}*a_{35}*a_{47} + 3a_{8}*a_{35}*a_{50} - 3a_{23}*a_{38}*a_{50} + a_{29}*a_{32}*a_{56} - a_{30}*a_{32}*a_{56} + a_{17}*a_{44} - a_{14}*a_{47} - a_{35}*a_{47} + a_{38}*a_{47} - a_{8}*a_{56} + a_{23}*a_{56}$
\item $
 3a_{17}*a_{32}*a_{44}^{2} - 6a_{14}*a_{32}*a_{44}*a_{47} + 3a_{32}*a_{35}*a_{44}*a_{47} + 3a_{14}*a_{30}*a_{47}^{2} - a_{29}*a_{35}*a_{47}^{2} - 2a_{30}*a_{35}*a_{47}^{2} - 3a_{8}*a_{17}*a_{44}*a_{50} - 3a_{17}*a_{32}*a_{44}*a_{50} + 3a_{14}*a_{32}*a_{47}*a_{50} - 3a_{8}*a_{35}*a_{47}*a_{50} + 3a_{23}*a_{38}*a_{47}*a_{50} + 3a_{8}*a_{17}*a_{50}^{2} - 3a_{17}*a_{23}*a_{44}*a_{53} + 3a_{8}*a_{35}*a_{47}*a_{53} - 3a_{8}*a_{38}*a_{47}*a_{53} - 3a_{17}*a_{23}*a_{50}*a_{53} + 6a_{8}*a_{32}*a_{44}*a_{56} - a_{29}*a_{32}*a_{47}*a_{56} + a_{30}*a_{32}*a_{47}*a_{56} - 3a_{8}^{2}*a_{50}*a_{56} + 3a_{8}*a_{23}*a_{50}*a_{56} - 3a_{8}*a_{32}*a_{50}*a_{56} + 3a_{8}^{2}*a_{53}*a_{56} - 3a_{8}*a_{23}*a_{53}*a_{56} + 3a_{8}*a_{32}*a_{53}*a_{56} - a_{35}*a_{47}^{2} + a_{38}*a_{47}^{2} + a_{17}*a_{47}*a_{50} + a_{17}*a_{47}*a_{53} - 2a_{32}*a_{47}*a_{56}$
\item $
 9a_{8}*a_{23}*a_{38}*a_{44}*a_{50} - 9a_{8}*a_{23}*a_{38}*a_{50}^{2} + 9a_{23}*a_{32}*a_{35}*a_{44}*a_{53} + 9a_{8}*a_{23}*a_{38}*a_{44}*a_{53} + 3a_{29}*a_{32}*a_{35}*a_{47}*a_{53} - 3a_{30}*a_{32}*a_{35}*a_{47}*a_{53} + 9a_{23}^{2}*a_{38}*a_{50}*a_{53} + 9a_{23}*a_{30}*a_{32}*a_{53}*a_{56} + 3a_{29}*a_{32}^{2}*a_{53}*a_{56} - 3a_{30}*a_{32}^{2}*a_{53}*a_{56} - 3a_{17}*a_{23}*a_{44}^{2} - 3a_{23}*a_{35}*a_{44}*a_{47} + 6a_{32}*a_{35}*a_{44}*a_{47} - 3a_{29}*a_{35}*a_{47}^{2} - 3a_{30}*a_{35}*a_{47}^{2} + 3a_{17}*a_{23}*a_{44}*a_{50} - 9a_{8}*a_{35}*a_{47}*a_{50} + 3a_{8}*a_{38}*a_{47}*a_{50} + 3a_{23}*a_{38}*a_{47}*a_{50} - 3a_{17}*a_{23}*a_{44}*a_{53} + 3a_{8}*a_{35}*a_{47}*a_{53} - 3a_{32}*a_{35}*a_{47}*a_{53} - 3a_{8}*a_{38}*a_{47}*a_{53} - 3a_{23}*a_{38}*a_{47}*a_{53} - 6a_{8}*a_{23}*a_{44}*a_{56} - 3a_{23}*a_{30}*a_{47}*a_{56} - 3a_{29}*a_{32}*a_{47}*a_{56} + 3a_{30}*a_{32}*a_{47}*a_{56} + 6a_{8}*a_{23}*a_{50}*a_{56} - 3a_{23}^{2}*a_{50}*a_{56} - 3a_{8}*a_{23}*a_{53}*a_{56} - 3a_{32}^{2}*a_{53}*a_{56} + a_{14}*a_{47}^{2} + 2a_{35}*a_{47}^{2} - a_{38}*a_{47}^{2} - a_{17}*a_{47}*a_{50} + a_{17}*a_{47}*a_{53} + 2a_{8}*a_{47}*a_{56} + a_{32}*a_{47}*a_{56}$
\item $
 9a_{23}*a_{32}*a_{35}*a_{44}^{2} + 9a_{23}*a_{30}*a_{35}*a_{44}*a_{47} + a_{29}^{2}*a_{35}*a_{47}^{2} + 4a_{29}*a_{30}*a_{35}*a_{47}^{2} - 5a_{30}^{2}*a_{35}*a_{47}^{2} - 18a_{23}*a_{32}*a_{35}*a_{44}*a_{50} - 6a_{29}*a_{32}*a_{35}*a_{47}*a_{50} + 6a_{30}*a_{32}*a_{35}*a_{47}*a_{50} + 27a_{23}*a_{30}*a_{38}*a_{47}*a_{50} + 9a_{29}*a_{32}*a_{38}*a_{47}*a_{50} - 9a_{30}*a_{32}*a_{38}*a_{47}*a_{50} + 18a_{8}*a_{23}*a_{35}*a_{50}^{2} - 9a_{8}*a_{23}*a_{38}*a_{50}^{2} - 9a_{23}^{2}*a_{38}*a_{50}^{2} - 9a_{23}^{2}*a_{35}*a_{44}*a_{53} + 9a_{23}*a_{32}*a_{35}*a_{44}*a_{53} + 9a_{8}*a_{23}*a_{38}*a_{44}*a_{53} + 9a_{23}*a_{30}*a_{35}*a_{47}*a_{53} + 6a_{29}*a_{32}*a_{35}*a_{47}*a_{53} - 6a_{30}*a_{32}*a_{35}*a_{47}*a_{53} - 9a_{23}^{2}*a_{35}*a_{50}*a_{53} + 9a_{23}^{2}*a_{38}*a_{50}*a_{53} + 9a_{23}*a_{30}*a_{32}*a_{44}*a_{56} + 3a_{29}*a_{32}^{2}*a_{44}*a_{56} - 3a_{30}*a_{32}^{2}*a_{44}*a_{56} + 9a_{23}*a_{30}^{2}*a_{47}*a_{56} + a_{29}^{2}*a_{32}*a_{47}*a_{56} + a_{29}*a_{30}*a_{32}*a_{47}*a_{56} - 2a_{30}^{2}*a_{32}*a_{47}*a_{56} + 9a_{23}^{2}*a_{30}*a_{50}*a_{56} - 27a_{23}*a_{30}*a_{32}*a_{50}*a_{56} - 9a_{29}*a_{32}^{2}*a_{50}*a_{56} + 9a_{30}*a_{32}^{2}*a_{50}*a_{56} - 9a_{23}^{2}*a_{30}*a_{53}*a_{56} + 18a_{23}*a_{30}*a_{32}*a_{53}*a_{56} + 6a_{29}*a_{32}^{2}*a_{53}*a_{56} - 6a_{30}*a_{32}^{2}*a_{53}*a_{56} - 3a_{17}*a_{23}*a_{44}^{2} - 9a_{23}*a_{35}*a_{44}*a_{47} - 3a_{32}*a_{35}*a_{44}*a_{47} + 9a_{23}*a_{38}*a_{44}*a_{47} - 3a_{29}*a_{35}*a_{47}^{2} + 3a_{29}*a_{38}*a_{47}^{2} - 3a_{30}*a_{38}*a_{47}^{2} + 6a_{17}*a_{23}*a_{44}*a_{50} - 3a_{8}*a_{35}*a_{47}*a_{50} - 9a_{23}*a_{35}*a_{47}*a_{50} + 6a_{32}*a_{35}*a_{47}*a_{50} + 6a_{8}*a_{38}*a_{47}*a_{50} + 6a_{23}*a_{38}*a_{47}*a_{50} - 9a_{32}*a_{38}*a_{47}*a_{50} - 3a_{17}*a_{23}*a_{44}*a_{53} + 3a_{8}*a_{35}*a_{47}*a_{53} + 3a_{23}*a_{35}*a_{47}*a_{53} - 6a_{32}*a_{35}*a_{47}*a_{53} - 3a_{8}*a_{38}*a_{47}*a_{53} - 3a_{23}*a_{38}*a_{47}*a_{53} - 3a_{8}*a_{23}*a_{44}*a_{56} + 3a_{23}^{2}*a_{44}*a_{56} - 3a_{32}^{2}*a_{44}*a_{56} - 12a_{23}*a_{30}*a_{47}*a_{56} - 4a_{29}*a_{32}*a_{47}*a_{56} + a_{30}*a_{32}*a_{47}*a_{56} - 3a_{23}*a_{32}*a_{50}*a_{56} + 9a_{32}^{2}*a_{50}*a_{56} - 3a_{8}*a_{23}*a_{53}*a_{56} + 3a_{23}^{2}*a_{53}*a_{56} + 3a_{23}*a_{32}*a_{53}*a_{56} - 6a_{32}^{2}*a_{53}*a_{56} + a_{14}*a_{47}^{2} + 4a_{35}*a_{47}^{2} - 4a_{38}*a_{47}^{2} - 2a_{17}*a_{47}*a_{50} + a_{17}*a_{47}*a_{53} - a_{8}*a_{47}*a_{56} + a_{23}*a_{47}*a_{56} + 4a_{32}*a_{47}*a_{56}$
\item $
 3a_{14}*a_{17}^{2}*a_{44}^{2} - 6a_{14}^{2}*a_{17}*a_{44}*a_{47} + 3a_{14}*a_{17}*a_{35}*a_{44}*a_{47} + 3a_{14}^{3}*a_{47}^{2} - 3a_{14}^{2}*a_{35}*a_{47}^{2} - 3a_{14}*a_{17}^{2}*a_{44}*a_{50} - 3a_{17}^{2}*a_{38}*a_{44}*a_{50} + 3a_{14}^{2}*a_{17}*a_{47}*a_{50} + 3a_{14}*a_{17}*a_{38}*a_{47}*a_{50} + 3a_{17}^{2}*a_{38}*a_{50}^{2} - 3a_{17}^{2}*a_{35}*a_{44}*a_{53} + 3a_{14}*a_{17}*a_{35}*a_{47}*a_{53} - 3a_{17}^{2}*a_{35}*a_{50}*a_{53} + 3a_{8}*a_{17}*a_{35}*a_{50}*a_{56} - 3a_{8}*a_{17}*a_{38}*a_{50}*a_{56} - 3a_{8}*a_{17}*a_{35}*a_{53}*a_{56} + 3a_{8}*a_{17}*a_{38}*a_{53}*a_{56} + 2a_{17}^{2}*a_{44}*a_{56} - 2a_{14}*a_{17}*a_{47}*a_{56} + a_{17}*a_{35}*a_{47}*a_{56} - a_{17}*a_{38}*a_{47}*a_{56} - a_{17}^{2}*a_{50}*a_{56} + a_{17}^{2}*a_{53}*a_{56}$
\item $
 3a_{29}^{2}*a_{35}*a_{38}*a_{44} + 3a_{29}*a_{30}*a_{35}*a_{38}*a_{44} - 6a_{30}^{2}*a_{35}*a_{38}*a_{44} + 3a_{29}^{2}*a_{35}*a_{38}*a_{50} + 12a_{29}*a_{30}*a_{35}*a_{38}*a_{50} + 12a_{30}^{2}*a_{35}*a_{38}*a_{50} - 6a_{29}^{2}*a_{38}^{2}*a_{50} + 3a_{29}*a_{30}*a_{38}^{2}*a_{50} + 3a_{30}^{2}*a_{38}^{2}*a_{50} - 3a_{29}^{2}*a_{35}^{2}*a_{53} - 12a_{29}*a_{30}*a_{35}^{2}*a_{53} - 12a_{30}^{2}*a_{35}^{2}*a_{53} - 3a_{29}^{2}*a_{35}*a_{38}*a_{53} - 3a_{29}*a_{30}*a_{35}*a_{38}*a_{53} + 6a_{30}^{2}*a_{35}*a_{38}*a_{53} - a_{29}^{3}*a_{35}*a_{56} - 3a_{29}^{2}*a_{30}*a_{35}*a_{56} + 4a_{30}^{3}*a_{35}*a_{56} - 2a_{29}^{3}*a_{38}*a_{56} - 3a_{29}^{2}*a_{30}*a_{38}*a_{56} + 3a_{29}*a_{30}^{2}*a_{38}*a_{56} + 2a_{30}^{3}*a_{38}*a_{56} + 3a_{29}*a_{35}^{2}*a_{44} + 6a_{30}*a_{35}^{2}*a_{44} - 6a_{29}*a_{35}*a_{38}*a_{44} - 3a_{30}*a_{35}*a_{38}*a_{44} + 3a_{29}*a_{38}^{2}*a_{44} - 3a_{30}*a_{38}^{2}*a_{44} - 9a_{14}*a_{30}*a_{35}*a_{50} - 9a_{14}*a_{30}*a_{38}*a_{50} - 12a_{29}*a_{35}*a_{38}*a_{50} - 6a_{30}*a_{35}*a_{38}*a_{50} + 12a_{29}*a_{38}^{2}*a_{50} + 6a_{30}*a_{38}^{2}*a_{50} + 18a_{14}*a_{30}*a_{35}*a_{53} + 3a_{29}*a_{35}^{2}*a_{53} + 6a_{30}*a_{35}^{2}*a_{53} - 9a_{30}*a_{35}*a_{38}*a_{53} - 3a_{29}*a_{38}^{2}*a_{53} + 3a_{30}*a_{38}^{2}*a_{53} + a_{29}^{2}*a_{35}*a_{56} + a_{29}*a_{30}*a_{35}*a_{56} - 2a_{30}^{2}*a_{35}*a_{56} + 2a_{29}^{2}*a_{38}*a_{56} - a_{29}*a_{30}*a_{38}*a_{56} - a_{30}^{2}*a_{38}*a_{56} - 3a_{14}*a_{35}*a_{44} - 3a_{35}^{2}*a_{44} + 3a_{14}*a_{38}*a_{44} + 9a_{35}*a_{38}*a_{44} - 6a_{38}^{2}*a_{44} + 3a_{14}^{2}*a_{50} + 6a_{14}*a_{35}*a_{50} - 6a_{14}*a_{38}*a_{50} + 3a_{35}*a_{38}*a_{50} - 3a_{38}^{2}*a_{50} - 3a_{14}^{2}*a_{53} - 3a_{14}*a_{35}*a_{53} - 6a_{35}^{2}*a_{53} + 3a_{14}*a_{38}*a_{53} + 3a_{35}*a_{38}*a_{53} + 3a_{38}^{2}*a_{53} - 2a_{29}*a_{35}*a_{56} - 4a_{30}*a_{35}*a_{56} + 2a_{29}*a_{38}*a_{56} + 4a_{30}*a_{38}*a_{56} + 2a_{35}*a_{56} - 2a_{38}*a_{56}$

\end{enumerate}

	\item $ N =2122$      
	(subset)
	\begin{enumerate}
\item $
a_{7} + 2a_{9} - 1$
\item $
 a_{8} - a_{9}$
\item $
 a_{28} + a_{29} + a_{30} - 1$
\item $
 a_{22} + 2a_{24} - 1$
\item $
 a_{23} - a_{24}$
\item $
 a_{31} + 2a_{33} - 1$
\item $
 a_{32} - a_{33}$
\item $
 a_{4} + 2a_{6} - 1$
\item $
 a_{5} - a_{6}$
\item $
 a_{19} + 2a_{21} - 1$
\item $
 a_{20} - a_{21}$
\item $
 a_{25} + 2a_{27} - 1$
\item $
 a_{26} - a_{27}$
\item $
 a_{1} + 2a_{3} - 1$
\item $
 a_{2} - a_{3}$
\item $
3a_{8}*a_{30} + a_{8} - a_{23} - a_{32}$
\item $
 a_{23}*a_{29} + 2a_{23}*a_{30} + a_{29}*a_{32} - a_{30}*a_{32} - a_{32}$
\item $
 3a_{8}*a_{29} - 2a_{8} + 2a_{23} - a_{32}$
\item $
 a_{5}*a_{29} + 2a_{5}*a_{30} - a_{5} + a_{20} - a_{26}$
\item $
 a_{2}*a_{29} + 2a_{2}*a_{30} - 3a_{5}*a_{32}$
\item $
 3a_{5}*a_{23} - 3a_{2}*a_{30} + 3a_{5}*a_{32} - a_{2}$
\item $
 3a_{8}*a_{20} - 3a_{8}*a_{26} + 3a_{5}*a_{32} - a_{2}$
\item $
 3a_{5}*a_{8} - a_{2}$
 \end{enumerate}

		\item $ N =2221$      
		(subset) 
		\begin{enumerate}
\item $
a_{13} + 2a_{15} - 1$
\item $
 a_{14} - a_{15}$
\item $
 a_{28} + a_{29} + a_{30} - 1$
\item $
 a_{34} + 2a_{36} - 1$
\item $
 a_{35} - a_{36}$
\item $
 a_{37} + 2a_{39} - 1$
\item $
 a_{38} - a_{39}$
\item $
 a_{10} + 2a_{12} - 1$
\item $
 a_{11} - a_{12}$
\item $
 a_{4} + 2a_{6} - 1$
\item $
 a_{5} - a_{6}$
\item $
 a_{19} + 2a_{21} - 1$
\item $
 a_{20} - a_{21}$
\item $
 a_{25} + 2a_{27} - 1$
\item $
 a_{26} - a_{27}$
\item $
a_{14}*a_{29} - a_{14}*a_{30} + a_{35} - a_{38}$
\item $
 a_{11}*a_{29} - a_{11}*a_{30} + 3a_{5}*a_{35} - 3a_{5}*a_{38}$
\item $
 a_{5}*a_{29} + 2a_{5}*a_{30} - a_{5} + a_{20} - a_{26}$
\item $
 3a_{14}*a_{20} - 3a_{14}*a_{26} + 3a_{11}*a_{30} - 3a_{5}*a_{35} + 3a_{5}*a_{38} - a_{11}$
\item $
 3a_{5}*a_{14} - a_{11}$
\item $
 a_{20}*a_{29}*a_{35} + 2a_{20}*a_{30}*a_{35} - a_{26}*a_{29}*a_{38} + a_{26}*a_{30}*a_{38} - a_{14}*a_{26} + a_{11}*a_{30} - a_{5}*a_{35} - a_{20}*a_{35} - a_{26}*a_{35} + 2a_{20}*a_{38}$
		\end{enumerate}

		\item $ N =2222$      
		(subset) 
		\begin{enumerate}
\item $
a_{16} + 2a_{18} - 1$
\item $
 a_{17} - a_{18}$
\item $
 a_{28} + a_{29} + a_{30} - 1$
\item $
 a_{13} + 2a_{15} - 1$
\item $
 a_{14} - a_{15}$
\item $
 a_{31} + 2a_{33} - 1$
\item $
 a_{32} - a_{33}$
\item $
 a_{7} + 2a_{9} - 1$
\item $
 a_{8} - a_{9}$
\item $
 a_{34} + 2a_{36} - 1$
\item $
 a_{35} - a_{36}$
\item $
 a_{37} + 2a_{39} - 1$
\item $
 a_{38} - a_{39}$
\item $
 a_{22} + 2a_{24} - 1$
\item $
 a_{23} - a_{24}$
\item $
 a_{10} + 2a_{12} - 1$
\item $
 a_{11} - a_{12}$
\item $
 a_{4} + 2a_{6} - 1$
\item $
 a_{5} - a_{6}$
\item $
 a_{19} + 2a_{21} - 1$
\item $
 a_{20} - a_{21}$
\item $
 a_{25} + 2a_{27} - 1$
\item $
 a_{26} - a_{27}$
\item $
 a_{1} + 2a_{3} - 1$
\item $
 a_{2} - a_{3}$
\item $
a_{17}*a_{30} - a_{14}*a_{32} - a_{8}*a_{35} + a_{8}*a_{38}$
\item $
 3a_{8}*a_{30} + a_{8} - a_{23} - a_{32}$
\item $
 a_{23}*a_{29} + 2a_{23}*a_{30} + a_{29}*a_{32} - a_{30}*a_{32} - a_{32}$
\item $
 a_{17}*a_{29} - a_{14}*a_{32} + 2a_{8}*a_{35} - 2a_{8}*a_{38}$
\item $
 a_{14}*a_{29} - a_{14}*a_{30} + a_{35} - a_{38}$
\item $
 a_{11}*a_{29} - a_{11}*a_{30} + 3a_{5}*a_{35} - 3a_{5}*a_{38}$
\item $
 3a_{8}*a_{29} - 2a_{8} + 2a_{23} - a_{32}$
\item $
 a_{5}*a_{29} + 2a_{5}*a_{30} - a_{5} + a_{20} - a_{26}$
\item $
 a_{2}*a_{29} + 2a_{2}*a_{30} - 3a_{5}*a_{32}$
\item $
 3a_{14}*a_{23} - 3a_{8}*a_{35} + 3a_{8}*a_{38} - a_{17}$
\item $
 3a_{5}*a_{23} - 3a_{2}*a_{30} + 3a_{5}*a_{32} - a_{2}$
\item $
 a_{17}*a_{20} - a_{11}*a_{23} - a_{17}*a_{26} + a_{11}*a_{32} + a_{2}*a_{35} - a_{2}*a_{38}$
\item $
 3a_{14}*a_{20} - 3a_{14}*a_{26} + 3a_{11}*a_{30} - 3a_{5}*a_{35} + 3a_{5}*a_{38} - a_{11}$
\item $
 3a_{8}*a_{20} - 3a_{8}*a_{26} + 3a_{5}*a_{32} - a_{2}$
\item $
 a_{5}*a_{17} - a_{11}*a_{23} + a_{2}*a_{35} - a_{2}*a_{38}$
\item $
 3a_{8}*a_{14} - a_{17}$
\item $
 3a_{5}*a_{14} - a_{11}$
\item $
 a_{2}*a_{14} - a_{11}*a_{23} + a_{2}*a_{35} - a_{2}*a_{38}$
\item $
 a_{8}*a_{11} - a_{11}*a_{23} + a_{2}*a_{35} - a_{2}*a_{38}$
\item $
 3a_{5}*a_{8} - a_{2}$
\item $
 a_{20}*a_{29}*a_{35} + 2a_{20}*a_{30}*a_{35} - a_{26}*a_{29}*a_{38} + a_{26}*a_{30}*a_{38} - a_{14}*a_{26} + a_{11}*a_{30} - a_{5}*a_{35} - a_{20}*a_{35} - a_{26}*a_{35} + 2a_{20}*a_{38}$
\item $
 6a_{8}*a_{26}*a_{35} - 3a_{5}*a_{32}*a_{35} - 3a_{20}*a_{32}*a_{35} - 3a_{8}*a_{26}*a_{38} - 3a_{23}*a_{26}*a_{38} + 6a_{5}*a_{32}*a_{38} + a_{17}*a_{26} - a_{11}*a_{32} + a_{2}*a_{35} - a_{2}*a_{38}$
		\end{enumerate}
\end{itemize}




\section{Algorithms}

\begin{algorithm}[H]
\caption{Inference of n-taxon phylogenetic network ($n > 8$) with phylogenetic invariants}
\label{algo:le8}
\Input{Table of estimated concordance factors; optional: the number of optimal networks ($m$) to return (default $m=5$)}
\Output{Top $m$ optimal networks with smallest invariant score}
scores $\gets$ an empty array\;
subnets $\gets$ an empty array\;
\For{$P_i$, a subset of 8 taxa}{
    $\text{scores}_i$,$\text{subnets}_i$ $\gets$ Algorithm \ref{algo8}(CF, $P_i$, m=2520); \tcc{Note total number of subnets (partitions for 8 taxa) = 2520}   
    append(scores,$\text{scores}_i$)\;
    append(subnets, $\text{subnets}_i$)\;
}
subnets\_sorted $\gets$ sort subnets in descending order according to its score\;
result $\gets$ an empty array\;

\For{i in 1:length(subnets\_sorted)}{
    miss\_taxa $\gets$ array of missing taxa in subnets\_sorted[i]\;
    \For{t in miss\_taxa}{
    \For{j in (i+1):length(subnets\_sorted)}{
        
        \If{t in $n_0$ of subnets\_sorted[j]}{
        add t to $n_0$ of subnets\_sorted[i]\;
        break\;
        }\ElseIf{t in $n_3$ of subnets\_sorted[j]}{
        add t to $n_3$ of subnets\_sorted[i]\;
        break\;
        }\ElseIf{t in $n_1$ of subnets\_sorted[j]}{
        \If{($n_2$ of subnets\_sorted[j] == $n_1$ of subnets\_sorted[i]) \& 
        ($n_1$ of subnets\_sorted[j] without t is in $n_2$ of subnets\_sorted[i])}{
        add t to $n_2$ of subnets\_sorted[i]\;
        break\;
        }\Else{
        add t to $n_1$ of subnets\_sorted[i]\;
        break\;
        }
        }\ElseIf{t in $n_2$ of subnets\_sorted[j]}{
        \If{($n_1$ of subnets\_sorted[j] == $n_2$ of subnets\_sorted[i]) \& 
        ($n_2$ of subnets\_sorted[j] without t is in $n_1$ of subnets\_sorted[i])}{
        add t to $n_1$ of subnets\_sorted[i]\;
        break\;
        }\Else{
        add t to $n_2$ of subnets\_sorted[i]\;
        break\;
        }
        }
        }
    
    }
    \If{subnets\_sorted[i] not in result}{
    append(result, subnets\_sorted[i])
    }
    \If{length(result)==m}{
        break\;
    }
}
\textbf{return} result\;
\end{algorithm}

\section{Simulation study}

\begin{table}[h!]
  \begin{center}
    \begin{tabular}{l|c}
      \toprule 
      \textbf{Method} & \textbf{Time (seconds)} \\
      Phylogenetic invariants (our method)    & 7.17 \\
      \texttt{SNaQ}          & 80.23 \\
      \texttt{PhyloNet ML}   & 84.34 \\
      \texttt{PhyloNet MPL}  & 17.99 \\
      \bottomrule 
    \end{tabular}
    \caption{Running times (in seconds) on the inference of network $N=2222$ from 100 true simulated gene trees by four network methods: 1) our phylogenetic invariants (top row); 2) \texttt{SNaQ} \cite{snaq}; 3) \texttt{PhyloNet ML} \cite{Yu2014}, and 4) \texttt{PhyloNet MPL} \cite{yu2015maximum}.}
    \label{sim-time}
  \end{center}
\end{table}

\begin{table}[h!]
  \begin{center}
    \begin{tabular}{c|c|c}
      \toprule 
      \textbf{Network} & \textbf{True CF} &
      \textbf{Noisy CF}\\
      2223 & 1 & 1 \\
      2232 & 1 & 1 (symmetric) \\
      2322 & 1& 2 \\
      3222 & 1& 1 \\
      2233 & 1 & 1 (symmetric) \\
      2323 &  1 & 2 (symmetric)\\
      3223 & 1 & 1 \\
      2332 & 1 & 2 (symmetric) \\
      3232 & 1 & 2 (symmetric) \\
      3322 & 1 & 1 (symmetric) \\
      \bottomrule 
    \end{tabular}
  \end{center}
  \caption{Rank of the true (or symmetric) network under the two types of simulations: true CFs (left column) and Gaussian-perturbed CFs (right column) for a level of noise of $\sigma=0.0005$. In all cases, either the true network or the symmetric networks are within the top 2 of networks identified by the method which proves that our algorithm for more than 8 taxa that builds on Algorithm \ref{algo8} works appropriately.}
  \label{table9-10}
\end{table}

\begin{figure}[h!]
\centering
\includegraphics[scale=0.65]{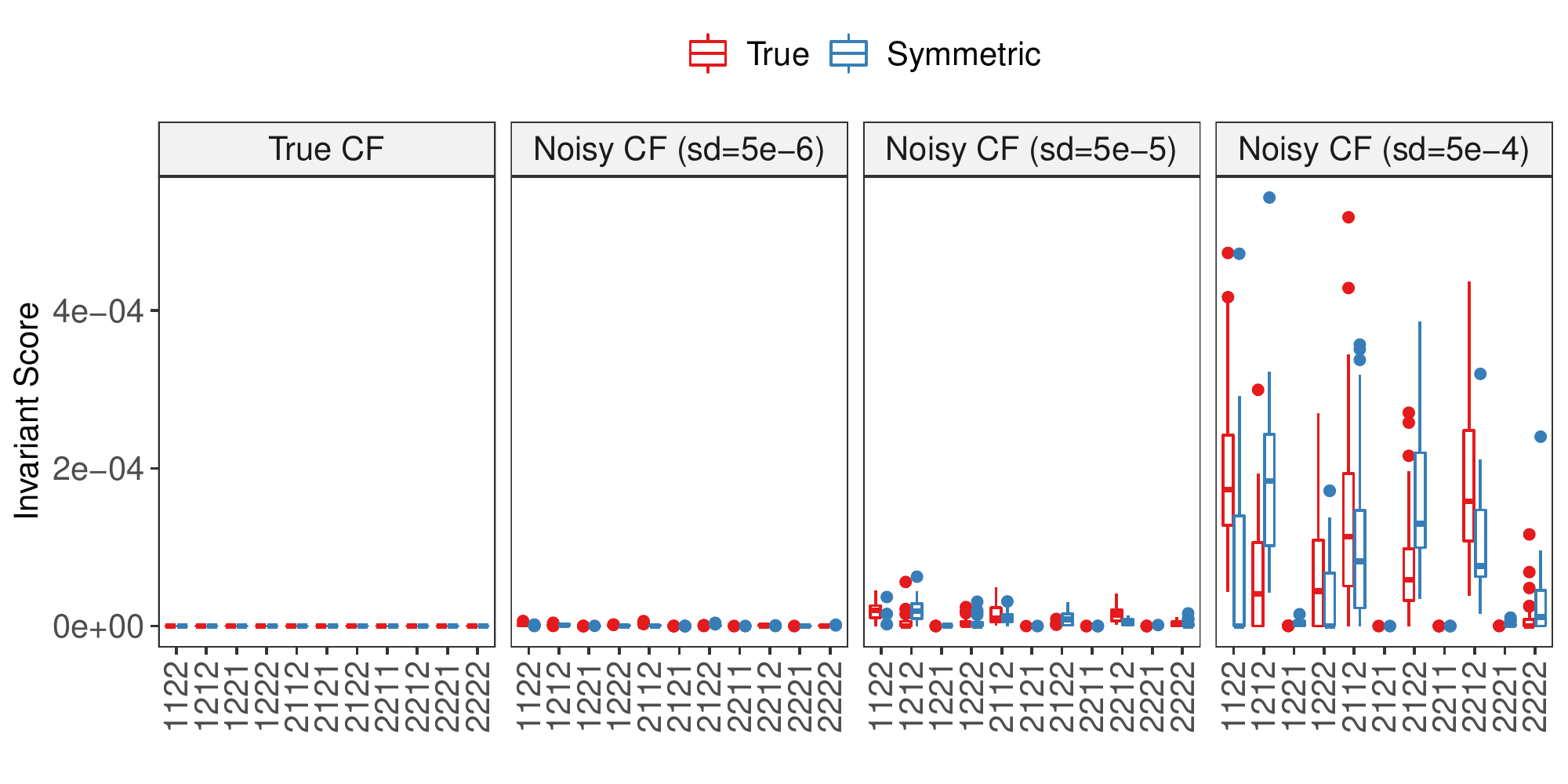}
\caption{Invariant score (y-axis) measured as $L_2$-norm of the phylogenetic invariants for each network (x-axis) for the true network (red) and its symmetric network (inverted clades $n_1$ and $n_2$, in blue) on the cases of true and Gaussian-perturbed CFs. Each panel corresponds to a type of simulation: using true concordance factors (left) and using concordance factors with added Gaussian noise (with increasing standard deviation for noise from left to right). Both the true and symmetric networks have invariant score close to zero, and are thus, easy to distinguish from wrong networks (whose invariant score is far from zero). As the noise increases, the invariant scores moves away from zero, but still within $10^{-4}$.}
\label{inv-score-truecf}
\end{figure}

\begin{figure}[h!]
\centering
\includegraphics[scale=0.65]{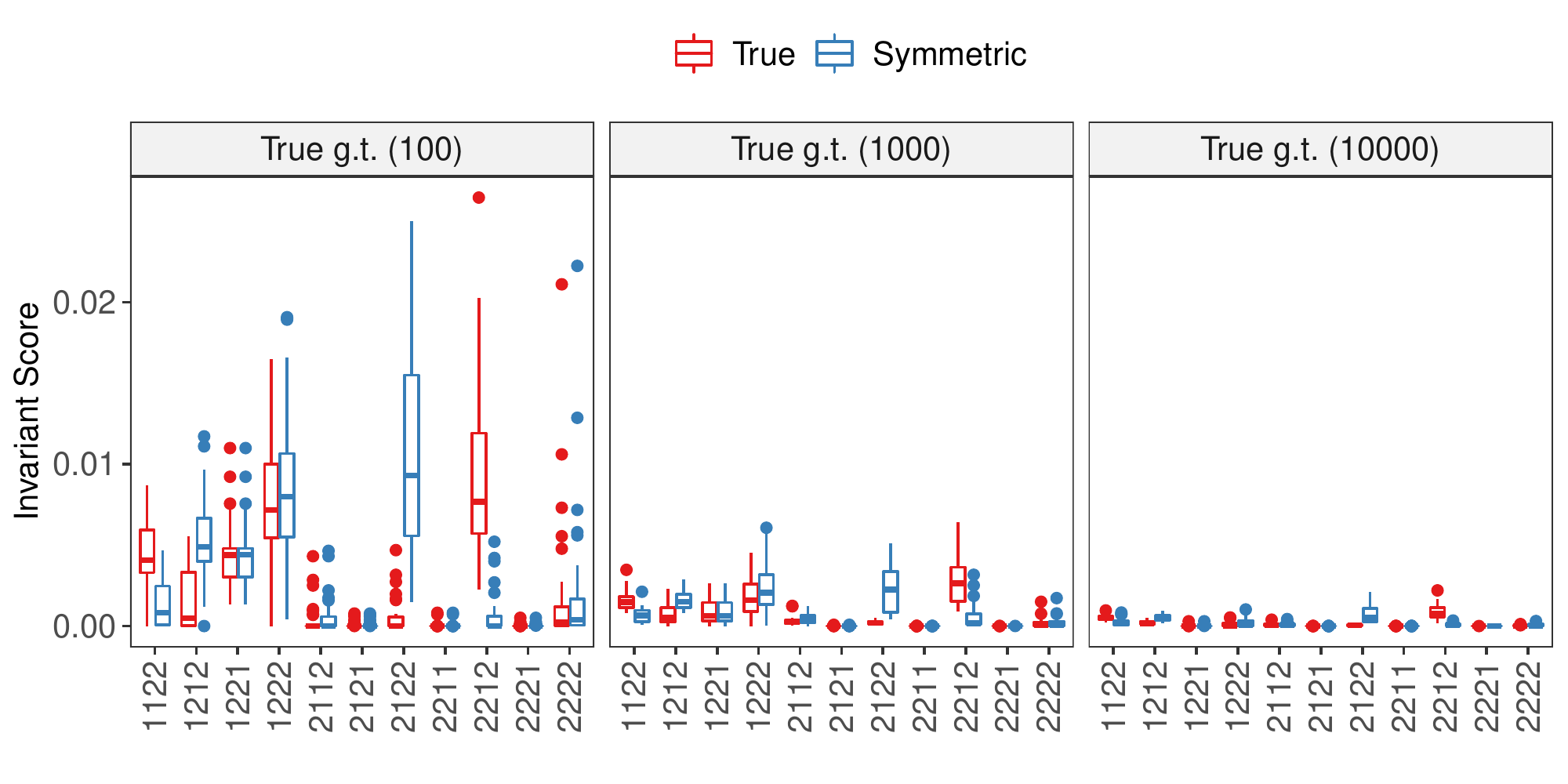}
\caption{Invariant score (y-axis) measured as $L_2$-norm of the phylogenetic invariants for each network (x-axis) for the true network (red) and its symmetric network (inverted clades $n_1$ and $n_2$, in blue) for the case of true simulated gene trees (``g.t."). Each panel corresponds to a number of simulated gene trees from 100 (left) to 10,000 (right). With true and perturbed concordance factors (top left and top right, respectively). As the number of gene trees increases, the invariant scores of the true and symmetric networks converge to zero, and they are thus, easy to distinguish from wrong networks (whose invariant score is far from zero).}
\label{inv-score-truegt}
\end{figure}

\begin{figure}[h!]
\centering
\includegraphics[scale=0.65]{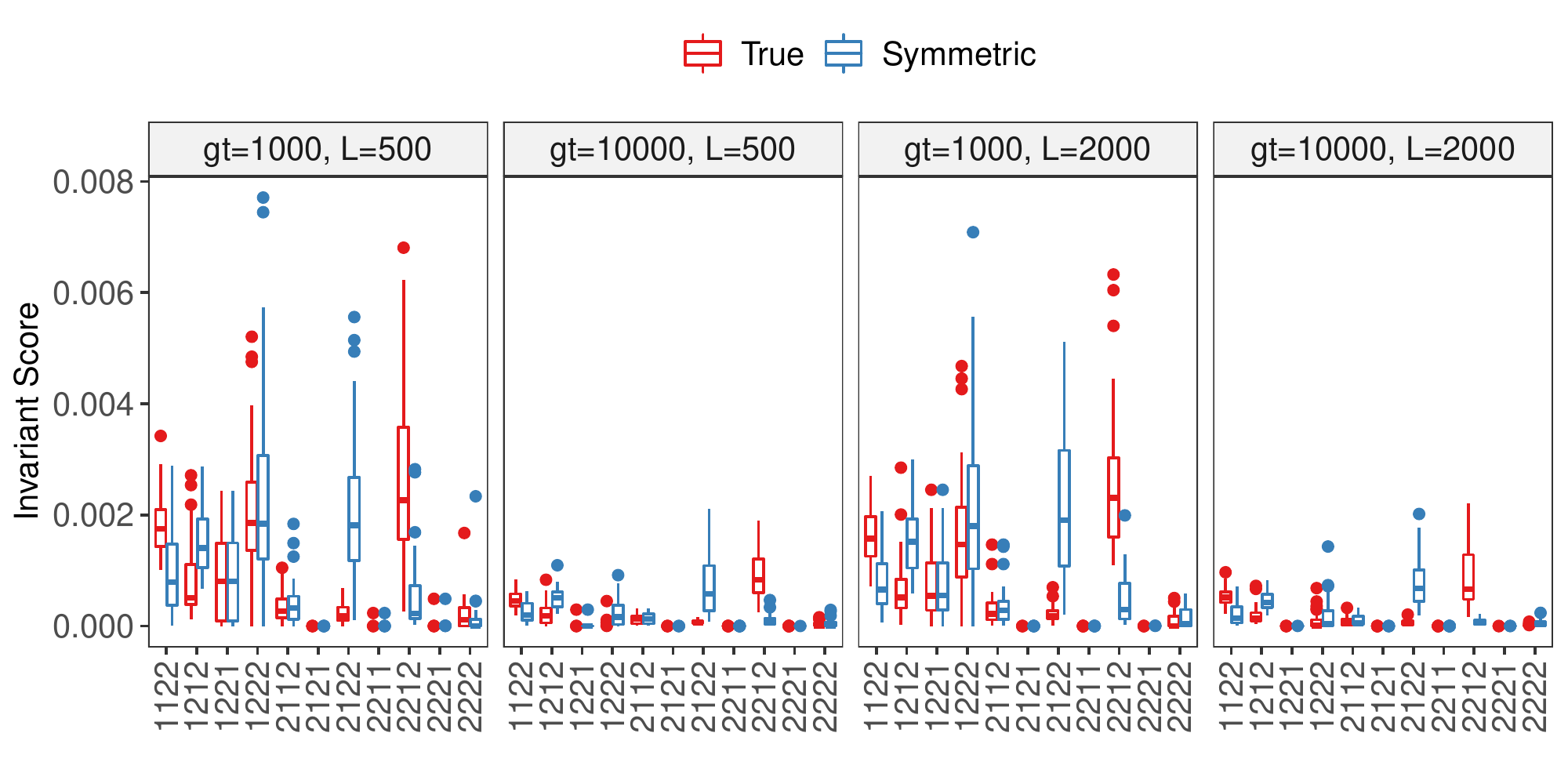}
\caption{Invariant score (y-axis) measured as $L_2$-norm of the phylogenetic invariants for each network (x-axis) for the true network (red) and its symmetric network (inverted clades $n_1$ and $n_2$, in blue) for the case of estimated gene trees. Each panel corresponds to a number of gene trees (g.t. from 1000 to 10,000) and sequence length ($L$ from 500 to 2000). As the number of gene trees increases, the invariant scores of the true and symmetric networks converge to zero, and they are thus, easy to distinguish from wrong networks (whose invariant score is far from zero).}
\label{inv-score-estgt}
\end{figure}

\end{document}